\documentclass[copyright,creativecommons]{eptcs} 
\providecommand{\event}{ICE 24} % Name of the event you are submitting to
\usepackage{underscore}           % Only needed if you use pdflatex.
 
\usepackage{breakurl}

\usepackage{wrapfig}

\usepackage{hyperref}
\usepackage{amsfonts}
\usepackage{amsmath}
\usepackage{amssymb}
\usepackage{amsthm}
\usepackage{amsbsy}
\usepackage{enumerate}
\usepackage{stackrel}
\usepackage{bm}
\usepackage{stmaryrd}
\usepackage{wasysym}
\usepackage{color}
\usepackage[utf8]{inputenc}
\usepackage{dashbox}
\usepackage[capitalise]{cleveref}
\usepackage{accents}
\crefformat{enumi}{condition~#2#1#3}
\crefname{fact}{Fact}{Facts}
\Crefname{fact}{Fact}{Facts}
\crefformat{fact}{Fact~#2#1#3}

\def\colorPtp{\color{blue}}
\def\colorNode{\color{cyan}}
\def\colorOp{\color{OliveGreen}}
\def\colorMsg{\color{BrickRed}}

\usepackage{amsmath}
\usepackage{amssymb}
\usepackage{mfirstuc}
\newcommand{\aG}{\mathsf{G}}
\newcommand{\gatedistancein}{3pt}
\newcommand{\gatedistanceinand}{2pt}
\newcommand{\gname}[1][i]{{\colorNode{\scriptstyle\textsf{#1}}}}
\usepackage{xifthen}        % for conditional commands
\usepackage{xstring}
\usepackage{xargs}
\usepackage[usenames,dvipsnames,svgnames,table]{xcolor} % must be loaded before tikz
\usepackage{tikz}

\usetikzlibrary{
  arrows,snakes,shapes,automata,backgrounds,petri,positioning,fit,calc,
  decorations.markings,decorations.text,shadows,fadings,patterns,
  decorations.pathreplacing,chains,spy
}

\tikzset{
  src/.style={draw,circle,fill=white,
    minimum size=2mm,
    inner sep=0pt
  },
  sink/.style={draw,circle,double,fill=white,
    minimum size=1.5mm,
    inner sep=0pt
  },
  node/.style={draw,circle,fill=black,
    minimum size=2mm,
    inner sep=0pt
  },
  source/.style={draw,circle,fill=white,
    minimum size=3mm,
    inner sep=0pt
  },
  sink/.style={draw,circle,double,fill=white,
    minimum size=3mm,
    inner sep=0pt
  },
  block/.style = {rectangle, draw=gray, align=center, fill=orange!25, rounded corners=0.1cm,
    minimum size=5mm, inner sep=2pt},
  prenode/.style = {minimum size=9pt,inner sep=2pt, font=\Large},
  bblock/.style = {rectangle, draw=blue!50, opacity=.5, line width=1pt, align=center, fill=white, rounded corners=0.1cm,
    minimum size=7mm, inner sep=2pt},
  prenode/.style = {minimum size=9pt,inner sep=2pt, font=\Large},
  agate/.style={draw, rectangle,
    minimum size=3mm,
    inner sep=0pt,
    fill=orange!25,
    postaction={path picture={% 
        \draw[red]
        ([yshift=\gatedistanceinand]path picture bounding box.south) --
        ([yshift=-\gatedistanceinand]path picture bounding box.north) ;}}
  },
  ogate/.style = {
    diamond, draw, fill=orange!25,
    minimum size=4mm,
    inner sep=0pt,
    postaction={path picture={% 
        \draw[red]
        ([yshift=\gatedistancein]path picture bounding box.south) -- ([yshift=-\gatedistancein]path picture bounding box.north)
        ([xshift=-\gatedistancein]path picture bounding box.east) -- ([xshift=\gatedistancein]path picture bounding box.west)
        ;}}},
  altogate/.style = {
    diamond, draw,
    minimum size=4mm,
    inner sep=0pt,
    postaction={path picture={% 
        \draw
        ([yshift=\gatedistancein]path picture bounding box.south) -- ([yshift=-\gatedistancein]path picture bounding box.north)
        ([xshift=-\gatedistancein]path picture bounding box.east) -- ([xshift=\gatedistancein]path picture bounding box.west)
        ;}}},
  altgate/.style={draw, rectangle,
    minimum size=3mm,
    inner sep=0pt,
    postaction={path picture={% 
        \draw
        ([yshift=\gatedistanceinand]path picture bounding box.south) --
        ([yshift=-\gatedistanceinand]path picture bounding box.north) ;}}},
  anygate/.style = {circle, draw, fill=white,
    minimum size=4mm,
    inner sep=0pt,
    postaction={path picture={% 
        \draw[black]
        ([xshift=-\gatedistancein,yshift=\gatedistancein]path picture bounding box.south east) --
        ([xshift=\gatedistancein,yshift=-\gatedistancein]path picture bounding box.north west)
        ([xshift=-\gatedistancein,yshift=-\gatedistancein]path picture bounding box.north east) --
        ([xshift=\gatedistancein,yshift=\gatedistancein]path picture bounding box.south west)
        ;}}
  },
  smallglobal/.style={
        node distance=1cm and 0.8cm, semithick, scale=0.8, every node/.style={transform shape}
  },
  elli/.style = {draw,densely dotted,-},
  line/.style = {draw,->, rounded corners=0.07cm,>=latex},
  nline/.style = {draw,semithick, ->},
  pline/.style = {draw,->,>=latex},
  node distance=1cm and 0.7cm,
  baseline=(current  bounding  box.center),
  local/.style={rectangle, draw, fill=\fillcolor, drop shadow,
    text centered, rounded corners, minimum height=5em
  },
  bigar/.style={
    draw,very thick, ->
  },
  process/.style={rectangle, draw=gray, fill=\fillcolor, drop shadow,
    text centered, minimum height=5em,text=gray
  },
  choreo/.style={rectangle, draw, fill=\fillcolor, drop shadow,
    text centered, rounded corners, minimum height=5em
  },
  mycfsm/.style={
        font=\footnotesize,
        initial where=above,
        ->,>=stealth,auto, node distance=1cm and 1cm,
        scale=1, every node/.style={transform shape},
        every state/.style=inner sep=2pt,
        baseline=(current  bounding  box.center)
  },
  machinecloud/.style={
    cloud, cloud puffs=10, cloud ignores aspect, minimum height=.1cm, minimum width=2cm, draw
  },
  fitting node/.style={
    inner sep=0pt,
    fill=none,
    draw=none,
    reset transform,
    fit={(\pgf@pathminx,\pgf@pathminy) (\pgf@pathmaxx,\pgf@pathmaxy)}
  },
  mypetri/.style={
    font=\footnotesize,
    baseline=(current  bounding  box.center)
  },
  silentrans/.style = {rectangle, draw=black, align=center, fill=black,
    minimum height=1pt,
    minimum width=15pt,
    inner sep=1.5pt
  },
  reset transform/.code={\pgftransformreset},
  tmtape/.style={draw,minimum size=1.2cm}
}

\newcommand{\p}{\ptp}
\newcommand{\q}{{\ptp[b]}}
\newcommand{\msg}[1][m]{\mathsf{\colorMsg{#1}}}
\newcommand{\ifempty}[3]{%
  \ifthenelse{\isempty{#1}}{#2}{#3}%
}
\newcommandx{\nmerge}[2][1={i},2={},usedefault=@]{
  \ifempty{#2}{
    \ifempty{#1}{\mu}{-\gname[{#1}]}
  }{-{#2}}
}
\newcommandx{\gint}[4][1=i,2=\ptp,3=\msg,4=\q,usedefault=@]{
  \ptp[{#2}] {\colorOp \xrightarrow{\scriptscriptstyle\gname[#1]}} \ptp[{#4}] \colon {\msg[{#3}]}
}
\newcommandx{\gout}[4][1=\gname,2=\ptp,3=\msg,4={\ptp[C]},usedefault=@]{
  \achan[{#2}][{#4}] {\colorOp {\colorOp{!}}} {\msg[{#3}]}
}
\newcommandx{\gin}[4][1=\gname,2=\ptp,3=\msg,4={\ptp[C]},usedefault=@]{
  \achan[{#2}][{#4}] {\colorOp {\colorOp{?}}} {\msg[{#3}]}
}
\newcommandx{\gseq}[3][1=i,2={\aG},3={\aG'},usedefault=@]{
  \gnode[{#1}][{#2} \gseqop {#3}]
}
\newcommandx{\gpar}[3][1=i,2={\aG},3={\aG'},usedefault=@]{
  \gnode[{#1}][\ifempty{#1}{{#2} \gparop {#3}}{({#2} \gparop {#3})}]
}
\newcommandx{\gcho}[3][1=i,2={\aG},3={\aG'},usedefault=@]{
  \gnode[{#1}][\ifempty{#1}{{#2} \gchoop {#3}}{\big({#2} \gchoop {#3}\big)}]
}
\newcommandx{\gchov}[3][1=i,2={\aG},3={\aG'},usedefault=@]{
  \gnode[{#1}][\left(
  \begin{array}l
    \ifempty{#1}{{#2} \\ \gchoop \\ {#3}}{\!\!{#2} \\ \gchoop \\ {#3}}
  \end{array}\right)
  ]
}
\newcommandx{\grec}[3][1=i,2={\aG},3={\p},usedefault=@]{
  \gnode[{#1}][\ifempty{#1}{\grecop {#2} \grecopp {#3}}{\big(\grecop {#2} \grecopp {#3}\big)}]
}
\newcommand{\ptp}[1][A]{
  \ensuremath{\mathtt{\colorPtp{%\capitalisewords
  {#1}}}}
}

\newcommandx{\mkint}[6][3=i,4=\p,5=\msg,6=\q,usedefault=@]{
  \node[bblock,{#1}] (#2) {$\gint[#3][#4][#5][#6]$};
}

\newcommandx{\mkgraph}[3][1=.5cm]{
  \node[source,above = #1 of {#2}] (src#2) {};
  \node[sink,below  = #1 of {#3}] (sink#3) {};
  \path[line] (src#2) -- (#2);
  \path[line] (#3) -- (sink#3);
}

\newcommandx{\mkloop}[4][1=.5,2=1.5]{ %it does insert an agate below and one above
  \node[ogate,above = #1 of {#3}] (entry#3) {};
  \pgfgetlastxy \xentry \yentry;
  \pgfmathtruncatemacro{\xentryrounded}{\xentry};
  \node[ogate,below  = #1 of {#4}] (exit#4) {};
  \pgfgetlastxy \xexit \yexit;
  \pgfmathtruncatemacro{\xexitrounded}{\xexit};
  \path[line] (entry#3) -- (#3);
  \path[line] (#4) -- (exit#4);
  \pgfmathsetmacro\tmpdiff{abs(\xentryrounded - \xexitrounded)}
  \path[line] (exit#4) -|  ($(exit#4)+(\tmpdiff,0)+(#2,0)$) |- (entry#3);
}

\newcommandx{\mklooptwo}[4][1=.5,2=1.5]{ %does not insert the a gate below
  \node[ogate,above = #1 of {#3}] (entry#3) {};
  \pgfgetlastxy \xentry \yentry;
  \pgfmathtruncatemacro{\xentryrounded}{\xentry};
  \path (#4);
   \pgfgetlastxy \xexit \yexit;
  \pgfmathtruncatemacro{\xexitrounded}{\xexit};
  \path[line] (entry#3) -- (#3);
  \pgfmathsetmacro\tmpdiff{abs(\xentryrounded - \xexitrounded)}
  \path[line] (#4) -|  ($(#4)+(\tmpdiff,0)+(#2,0)$) |- (entry#3);
}

\newcommandx{\mklooptwobelow}[4][1=.5,2=1.5]{ %does not insert the a gate below, add extra line down
  \node[ogate,above = #1 of {#3}] (entry#3) {};
  \pgfgetlastxy \xentry \yentry;
  \pgfmathtruncatemacro{\xentryrounded}{\xentry};
  \path (#4);
   \pgfgetlastxy \xexit \yexit;
  \pgfmathtruncatemacro{\xexitrounded}{\xexit};
  \path[line] (entry#3) -- (#3);
  \pgfmathsetmacro\tmpdiff{abs(\xentryrounded - \xexitrounded)}
  \path[line] (#4)  |- ($(#4)+(0,-0.5)$) -|  ($(#4)+(\tmpdiff,0)+(#2,0)$) |- (entry#3);
}

\newcommandx{\mklooponetwo}[4][1=.5,2=1.5]{ %does not insert the a gate below
  \path (#3);
  \pgfgetlastxy \xentry \yentry;
  \pgfmathtruncatemacro{\xentryrounded}{\xentry};
  \path (#4);
   \pgfgetlastxy \xexit \yexit;
  \pgfmathtruncatemacro{\xexitrounded}{\xexit};
  \path[line] (entry#3) -- (#3);
  \pgfmathsetmacro\tmpdiff{abs(\xentryrounded - \xexitrounded)}
  \path[line] (#4) -|  ($(#4)+(\tmpdiff,0)+(#2,0)$) |- (entry#3);
}

\newcommandx{\mkfork}[4][2=gatenode,3=i,4=.6,usedefault=@]{
  \mkgatebegin{#1}[{\gname[#3]}][agate][#4]{#2}
}

\newcommandx{\mkbranch}[4][2=gatenode,3=i,4=.6,usedefault=@]{
  \mkgatebegin{#1}[{\gname[#3]}][ogate][#4]{#2}
}

\newcommandx{\mkgatebegin}[5][2={},3=ogate,4=.5]{
  \coordinate (gatecord) at (0,0);
  \foreach \n [count=\i] in {#1}{
    \pgfgetlastxy \xc \yc;
    \path (\n);
    \pgfgetlastxy \xn \yn;
    \coordinate (gatecord) at ($(gatecord) + (\xn,0)$);
    \coordinate (gatecord) at ($1/\i*(gatecord)$);
    \ifdim \yn < \yc
    \node (max) at (0,\yc) {};
    \else
    \node (max) at (0,\yn) {};
    \fi
  }
  \coordinate (gatecord) at ($(gatecord) + (0,#4) + (max)$);
  \node[#3,label={below:$#2$}] (#5) at (gatecord) {};
  \pgfgetlastxy{\xgate}{\ygate};
  \pgfmathtruncatemacro{\xgateround}{\xgate};
  \StrCount{#1,}{,}[\l] % from package xxstring
  \ifnum \l < 2 {\errmessage{#1 argument should be a comma-separated list of lenght >= 2}}
  \else{
    \foreach \n in {#1}{
      \path (\n);
      \pgfgetlastxy{\xnode}{\ynode};
      \pgfmathtruncatemacro{\xnround}{\xnode};
      \pgfmathsetmacro\tmpdiff{abs(\xnround - \xgateround)}
      \ifdim \tmpdiff pt > 1 pt \path[line] (#5) -| (\n);
      \else
        \path[line] (#5) -- (\n);
      \fi
    }
  }
  \fi
}

\newcommandx{\mkmerge}[4][2=gatenode,3=i,4=0,usedefault=@]{\mkgateend{#1}[{\ifempty{#3}{}{\nmerge[#3]}}][ogate][#4]{#2}}

\newcommandx{\mkjoin}[4][2=gatenode,3=i,4=0,usedefault=@]{\mkgateend{#1}[{\ifempty{#3}{}{\nmerge[#3]}}][agate][#4]{#2}}

\newcommandx{\mkgateend}[5][2={},3=ogate,4=.5]{
  \coordinate (gatecord) at (0,0);
  \foreach \n [count=\i] in {#1}{
    \pgfgetlastxy \xc \yc;
    \path (\n);
    \pgfgetlastxy \xn \yn;
    \coordinate (gatecord) at ($(gatecord) + (\xn,0)$);
    \coordinate (gatecord) at ($1/\i*(gatecord)$);
    \ifdim \yn > \yc
    \node (min) at (0,\yc) {};
    \else
    \node (min) at (0,\yn) {};
    \fi
  }
  \coordinate (gatecord) at ($(gatecord) - (0,#4) + (min)$);
  \node[#3,label={above:$#2$}] (#5) at (gatecord) {};
  \pgfgetlastxy{\xgate}{\ygate};
  \pgfmathtruncatemacro{\xgateround}{\xgate};
  \StrCount{#1,}{,}[\l] % from package xxstring
  \ifnum \l < 2 {\errmessage{#1 argument should be a comma-separated list of lenght >= 2}}
  \else{
    \foreach \n in {#1}{
      \path (\n);
      \pgfgetlastxy{\xnode}{\ynode};
      \pgfmathtruncatemacro{\xnround}{\xnode};
      \pgfmathsetmacro\tmpdiff{abs(\xnround - \xgateround)}
      \ifdim \tmpdiff pt > 1 pt \path[line] (\n) |- (#5);
      \else
        \path[line] (\n) -- (#5);
      \fi
    }
  }
  \fi
}

\newtheorem{definition}{Definition}[section]

\newtheorem{proposition}[definition]{Proposition}
\newtheorem{theorem}[definition]{Theorem}

\newtheorem{remark}[definition]{Remark}
\newtheorem{example}[definition]{Example}

\DeclareMathAlphabet{\mathpzc}{OT1}{pzc}{m}{it}

\renewcommand*{\dot}[1]{%
  \accentset{\mbox{\large\bfseries .}}{#1}}

\newcommand{\Comment}[1]{ }

\def\Pred[#1]{~[\,#1\,]}

\newcommand{\RS}{\mathsf{RC}}

\newcommand{\gts}{{\looparrowleft}}

\newcommand{\inn}[1]{\mathsf{in}(#1)}
\newcommand{\Msgs}[1]{\mathsf{Msg}(#1)}
\newcommand{\outt}[1]{\mathsf{out}(#1)}
\newcommand{\cm}{\text{\sc cm}}

  \def\finex{{\unskip\nobreak\hfil\penalty50\hskip1em\null\nobreak\hfil{\Large $\diamond$}\parfillskip=0pt\finalhyphendemerits=0\endgraf}}
  
\newcommand{\cp}{\mathbb{K}}
\newcommand{\MC}{\mathcal{M}\!\mathcal{C}\!}

\newcommand{\HH}{{\ptp[h]}}
\newcommand{\hh}{\HH}
\newcommand{\kk}{\KK}
\newcommand{\pv}{{\ptp[v]}}
\newcommand{\pw}{{\ptp[w]}}
\newcommand{\KK}{{\ptp[k]}}

\newcommand{\roles}{\mathbf{P}}

\newcommand{\IS}[2]{\mathsf{L}\hspace{-1pt}\mathsf{C}\hspace{-1pt}\mathsf{P}\hspace{-1pt}\mathsf{S}(#1,#2)}

\newcommand{\restrict}[2]{{#1}_{\mid_{\mathbf{#2}}}}

\newcommand{\ttp}{{\ptp[p]}}
\newcommand{\ttq}{{\ptp[q]}}
\newcommand{\ttr}{{\ptp[r]}}

\newcommand{\tts}{{\ptp[s]}}

\newcommand{\ttu}{{\ptp[u]}}
\newcommand{\ttx}{{\ptp[x]}}

\newcommand{\ttv}{{\ptp[v]}}
\newcommand{\ttw}{{\ptp[w]}}

\newcommand{\elle}{\mathit{l}}

\newcommand{\Set}[1]{\{#1\}}

\renewcommand{\implies}{~\Longrightarrow~ }

\newcommand{\trans}[2][{}]{\,\xrightarrow{#2}_{#1}\,}
\newcommand{\lts}[1]{\trans{#1}}

\newcommand{\notlts}[1]{\stackrel{#1}{\;\;\not\!\!\longrightarrow}}

\newcommandx{\achan}[2][1=A,2=B,usedefault=@]{{\ptp[#1]\,\ptp[#2]}}

\newcommandx{\outop}[2][1=\gname,2={}]{{\colorOp{!}}^{{#1}{#2}}}
\newcommandx{\inop}[2][1=\gname,2={}]{{\colorOp{?}}^{{#1}{#2}}}

\newcommandx{\aout}[5][1={\p},2={\q},3={},4=m,5={},usedefault=@]{
  \achan[#1][#2] \outop[{#3}] {\msg[#4]}{#5}
}
\newcommandx{\ain}[5][1={\p},2={\q},3={},4=m,5={},usedefault=@]{
  \achan[#1][#2] \inop[{#3}] {\msg[#4]}{#5}
}

\tikzset{
  cnode/.style={
    shape=circle,
    minimum size = 0mm,
    inner sep = 1pt,
    font=\tiny,
    draw
  },
  carrow/.style={
    ->,
    shorten >=1pt,
    >=stealth',
    auto,
    draw,
    sloped
  }
}

\tikzset{
  src/.style={draw,circle,fill=white,
    minimum size=2mm,
    inner sep=0pt
  },
  sink/.style={draw,circle,double,fill=white,
    minimum size=1.5mm,
    inner sep=0pt
  },
  node/.style={draw,circle,fill=black,
    minimum size=2mm,
    inner sep=0pt
  },
  source/.style={draw,circle,fill=white,
    minimum size=3mm,
    inner sep=0pt
  },
  sink/.style={draw,circle,double,fill=white,
    minimum size=3mm,
    inner sep=0pt
  },
  block/.style = {rectangle, draw=gray, align=center, fill=orange!25, rounded corners=0.1cm,
    minimum size=5mm, inner sep=2pt},
  prenode/.style = {minimum size=9pt,inner sep=2pt, font=\Large},
  bblock/.style = {rectangle, draw=blue!50, opacity=.7, line width=.5pt, align=center, fill=white, rounded corners=0.1cm,
    minimum size=4mm, inner sep=1pt},
  prenode/.style = {minimum size=9pt,inner sep=2pt, font=\Large},
  agate/.style={draw, rectangle,
    minimum size=3mm,
    inner sep=0pt,
    fill=orange!25,
    label={[red]center:$\mid$}
  },
  ogate/.style = {
    diamond, draw, fill=orange!25,
    minimum size=4mm,
    inner sep=0pt,
    label={[red]center:$+$}
  },
  lgate/.style = {
    diamond, draw, fill=orange!25,
    minimum size=4mm,
    inner sep=0pt,
    label={[red]center:$\circlearrowleft$}
    },
  altogate/.style = {
    diamond, draw,
    minimum size=4mm,
    inner sep=0pt,
    postaction={path picture={% 
        \draw
        ([yshift=\gatedistancein]path picture bounding box.south) -- ([yshift=-\gatedistancein]path picture bounding box.north)
        ([xshift=-\gatedistancein]path picture bounding box.east) -- ([xshift=\gatedistancein]path picture bounding box.west)
        ;}}},
  altgate/.style={draw, rectangle,
    minimum size=3mm,
    inner sep=0pt,
    postaction={path picture={% 
        \draw
        ([yshift=\gatedistanceinand]path picture bounding box.south) --
        ([yshift=-\gatedistanceinand]path picture bounding box.north) ;}}},
  anygate/.style = {circle, draw, fill=white,
    minimum size=4mm,
    inner sep=0pt,
    postaction={path picture={% 
        \draw[black]
        ([xshift=-\gatedistancein,yshift=\gatedistancein]path picture bounding box.south east) --
        ([xshift=\gatedistancein,yshift=-\gatedistancein]path picture bounding box.north west)
        ([xshift=-\gatedistancein,yshift=-\gatedistancein]path picture bounding box.north east) --
        ([xshift=\gatedistancein,yshift=\gatedistancein]path picture bounding box.south west)
        ;}}
  },
  smallglobal/.style={
        node distance=1cm and 0.8cm, semithick, scale=0.8, every node/.style={transform shape}
  },
  elli/.style = {draw,densely dotted,-},
  line/.style = {draw,->, rounded corners=0.07cm,>=latex},
  nline/.style = {draw,semithick, ->},
  pline/.style = {draw,->,>=latex},
  node distance=1cm and 0.7cm,
  baseline=(current  bounding  box.center),
  local/.style={rectangle, draw, fill=\fillcolor, drop shadow,
    text centered, rounded corners, minimum height=5em
  },
  bigar/.style={
    draw,very thick, ->
  },
  process/.style={rectangle, draw=gray, fill=\fillcolor, drop shadow,
    text centered, minimum height=5em,text=gray
  },
  choreo/.style={rectangle, draw, fill=\fillcolor, drop shadow,
    text centered, rounded corners, minimum height=5em
  },
  mycfsm/.style={
        font=\footnotesize,
        initial where=above,
        ->,>=stealth,auto,
		  node distance=1.9cm,
        scale=.85,
		  every node/.style={transform shape},
        every state/.style={cnode, inner sep=1pt, transform shape},
		  every edge/.style={carrow},
        baseline=(current  bounding  box.center),
        initial text={}
  },
  machinecloud/.style={
    cloud, cloud puffs=10, cloud ignores aspect, minimum height=.1cm, minimum width=2cm, draw
  },
  fitting node/.style={
    inner sep=0pt,
    fill=none,
    draw=none,
    reset transform,
    fit={(\pgf@pathminx,\pgf@pathminy) (\pgf@pathmaxx,\pgf@pathmaxy)}
  },
  mypetri/.style={
    font=\footnotesize,
    baseline=(current  bounding  box.center)
  },
  silentrans/.style = {rectangle, draw=black, align=center, fill=black,
    minimum height=1pt,
    minimum width=15pt,
    inner sep=1.5pt
  },
  reset transform/.code={\pgftransformreset},
  tmtape/.style={draw,minimum size=1.2cm}
}

\def \bmr {\begin{color}{red}} 
\def \emr {\end{color}}
\def \bfr {\begin{color}{Fuchsia}} 
\def \efr {\end{color}}
\def \bmc {\begin{color}{magenta}Mariangiola: } 
\def \emc {\end{color}}
\def \bfc {\begin{color}{brown}Franco: } 
\def \efc {\end{color}}
\def \brc {\begin{color}{blue}Rolf: } 
\def \erc {\end{color}}
\def \brr {\begin{color}{olive}} 
\def \err {\end{color}}

\author{Franco Barbanera\thanks{ Partially supported by 
Project “National Center for HPC, Big Data e Quantum Computing”,  Programma M4C2, Investimento 1.3.}
\institute{Dipartimento di Matematica e Informatica\\
University of Catania}
\email{franco.barbanera@unict.it}
\and 
Rolf Hennicker
\institute{Institute for Informatics\\
LMU Munich}
\email{hennicke@pst.ifi.lmu.de}
}

\begin{document}

%\title{Multicomposition
% of Systems \\ of Communicating Finite State Machines
%}

\title{Safe Composition of Systems \\ of Communicating Finite State Machines
}

\def\titlerunning{Composition of CFSM Systems}
\def\authorrunning{F.\,Barbanera \& R.\,Hennicker}

\maketitle

\begin{abstract}
%OLD ABSTRACT
%The {\em Participants-as-Interfaces\/} (PaI) approach to system composition consists, roughly,
%in 
%choosing one partecipant per system and replacing them 
%by forwarders.
%Such an approach, for what concerns systems of Communicating Finite State Machines (CFSMs),
%has been exploited in the literature for binary composition only.
%We here consider the case of multiple system composition, when forwarders are not uniquely
%determined and their interactions depend on specific {\em connection policies\/}.
%We represent connection policies as CFSM systems and prove that a bunch of relevant 
%communication properties (deadlock-freeness, reception-error-freeness, etc.) are preserved
%by {\em PaI multicomposition\/}, with the proviso that also the used connection policy does enjoy
%the communication property taken into account.

The {\em Participants-as-Interfaces\/} (PaI) approach to system composition 
suggests that participants of a system may be viewed as interfaces.
Given a set of systems,
% of communicating finite state machines (CFSMs),  
one participant per system is chosen to play the role of an interface. When systems are composed, the %each CFSM
% of an 
interface participants are replaced by  \emph{gateways} which communicate to each other
%gateways of other systems 
by forwarding messages.
The PaI-approach for 
systems of asynchronous  communicating finite state machines (CFSMs)
has been exploited in the literature for binary composition only, with
a  (necessarily)  unique forwarding policy.
In this paper we consider the case of multiple system composition
when  forwarding gateways  are not uniquely determined and
their interactions depend on specific {\em connection policies\/}  complying with a
{\em connection model\/}.
We represent connection policies as CFSM systems and prove that a bunch of relevant 
communication properties (deadlock-freeness, reception-error-freeness, etc.) are preserved by {\em PaI multicomposition\/}, with the proviso that also the used connection policy does enjoy the communication property taken into account.
\end{abstract}

%!TEX root = Main-asynchCFSM-multicomp.tex
\section{Introduction}
\label{sec:Intro}
Concurrent/Distributed systems are  hardly -- especially nowadays -- 
stand-alone entities. They are part of 
``jigsaws'' never completely  finished.
Either in their design phase or after their deployment, they should be considered
as {\em open} and  ready for interaction with their environment, and hence with other systems. 
The possibility of extending and improving their functional and communication capabilities
by composing them
with other systems is also a crucial means %feature
against their obsolescence.
Compositional mechanisms and techniques are consequently an important subject for investigation.
As mentioned in \cite{BDGY23}, system composition investigations should focus on three relevant features
of these mechanisms/techniques:
\begin{itemize}
\item
{\em Conservativity:} 
They should alter as little as possible the single systems we compose.
\item
{\em Flexibility:}
They should not be embedded into the systems we compose, i.e.\ they should be 
``system independent''. 
In particular, they should allow to consider {\bf any} system as potentially {\bf open}.
%Even further, they should possibly allow to consider {\bf any} system as potentially {\bf open}.
\item
{\em Safety:}
Relevant properties of the single systems should not be ``broken'' by composition.
\end{itemize}

\begin{figure}[t]
    \centering{\small
    \vspace{-14mm}
    $
    \begin{array}{@{\hspace{-10mm}}c@{\hspace{-10mm}}}
 \begin{tikzpicture}[node distance=1.5cm,scale=1]
        \node (square-h) [draw,minimum width=0.8cm,minimum height=0.8cm] {\large $\hh$};
        \node [state] (h-a) [above of = square-h, draw=none] {};
        \node [state] (h-c) [left of = square-h, draw=none] {};
        \draw [-stealth] (h-a) --  node {$\msg[a]$} (square-h);
        %\draw [-stealth] (square-h) --  node {$\msg[c]$} (h-c);
 \end{tikzpicture}
\hspace{8mm}
\begin{array}{c}
\\[10mm]
| \\
| \\
| \\
|
\end{array}
 \hspace{8mm}
 \begin{tikzpicture}[node distance=1.5cm,scale=1]
        \node (square-k) [draw,minimum width=0.8cm,minimum height=0.8cm] {\large $\KK$};
        \node [state] (k-b) [above of = square-k, draw=none] {};
        \node [state] (k-a) [right of = square-k, draw=none] {};
        %\draw [-stealth] (k-b) --  node {$\msg[b]$} (square-k);
        \draw [-stealth] (square-k) --  node {$\msg[a]$} (k-a);
 \end{tikzpicture}
  \end{array}
  \qquad
  \begin{array}{c}
  \\[8mm]
  \text{becomes}
  \end{array}
  \qquad
  \begin{array}{c}
  \\[6mm]
  \begin{tikzpicture}[node distance=1.5cm,scale=1]
        \node (square-h) [draw,minimum width=0.8cm,minimum height=0.8cm] {\large $\hh$};
        \draw [-stealth] (h-a) --  node {$\msg[a]$} (square-h);
        \node (square-k) [draw,minimum width=0.8cm,minimum height=0.8cm, right of = square-h, xshift=5mm] {\large $\KK$};
        \node [state] (k-a) [right of = square-k, draw=none] {};
        \draw [-stealth] (square-k) --  node {$\msg[a]$} (k-a);
        \draw (0,0.4)[dotted,thick]  --  (0.4,0); % carrying a inside h  
        \draw [-stealth] (0.4,0)  -- node {$\msg[a]$} (1.6,0); % carrying a from h to k
        \draw (1.6,0)[dotted,thick]  --  (2.4,0); % carrying a inside k
 \end{tikzpicture}
 \end{array}
 $\vspace{-2mm}
 \caption{\label{fig:bincomp} The PaI idea for binary composition}
 }
\end{figure}
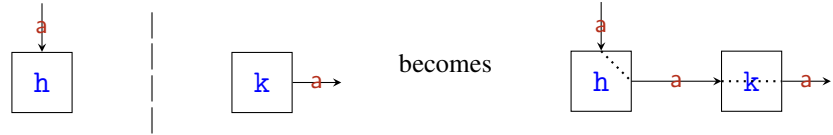

A fairly general and abstract approach to binary composition of systems 
was proposed in~\cite{BdLH19} and dubbed afterwards {\em Participants-as-Interfaces} (PaI). 
Roughly, the composition is achieved by transforming two selected participants  -- one per system,
say $\hh$ and $\kk$, -- 
into coupled forwarders (gateways), provided the participants exhibit ``compatible'' behaviours. 
 The graphics in~\cref{fig:bincomp} illustrates the PaI idea for the binary case.
If interface participant $\HH$ of the first system $S_1$
can receive a message $\msg[a]$  from some participant of $S_1$ and interface participant $\KK$
 of the second system $S_2$ can send $\msg[a]$ to some participant of $S_2$,  then the gateway replacing the first interface (also called $\HH$) will forward the received message to the gateway for $\KK$. 
How PaI works for multicomposition of systems will be illustrated
in~\cref{sec:pai-multicomp}.
 It is worth remarking that the PaI approach to system composition does not expect any particular condition to be satisfied by a single participant in order to be used as an interface. 
 
 \smallskip
{\em Conservativity} as well as {\em flexibility}
are definitely features of the PaI composition  idea.
 Conservativity holds since all participants not acting as interfaces remain untouched and flexibility holds since, in principle, any participant can play the role of an interface. This fact is independent of the concrete formalism used for protocol descriptions and system designs/implementations. 
{\em Safety}, instead, can be checked  only once we take into account a
specific formalism. 
Such checks were carried out in a number of papers where two relevant formalisms 
 for the description and verification of concurrent communicating systems were considered:
  MultiParty Session Types  (MPST)~\cite{HYC08,Honda2016}
and Communicating Finite State Machines  (CFSM)~\cite{BZ83}.
 %a structured one, MultiParty Session Types  (MPST)~\cite{HYC08,Honda2016},
Safety of the binary PaI approach was investigated for MPST
in~\cite{BDLT21}, where a synchronous communication model was considered.
The PaI approach to multicomposition  for MPST  has been exploited in~\cite{BDGY23,BDL22},
 again for synchronous communications. 
In particular, in~\cite{BDL22}, a restricted  notion of multiple connections in a client-server setting has been considered.
For the synchronous MPST formalism used in those papers, 
%and for synchronous communications, 
PaI proved to be safe.
The binary PaI approach for safety in systems of (standard) asynchronous CFSMs was taken into account in~\cite{BdLH19}, whereas 
safety of PaI for a synchronous version of the CFSM formalism was 
investigated in~\cite{BLT20,BLP22b,BLT23},  again for binary composition. 
%\brc Do~\cite{BLT20,BLP22b,BLT23} consider multicomposition or only binary composition?\erc

%In the mentioned papers, compatibility is essentially a bisimulation.
\medskip
\noindent
\emph{Contributions.}
 In the present paper we investigate safety of PaI multicomposition for the asynchronous formalism of CFSMs. For this purpose we reuse the PaI multicomposition idea of~\cite{BDGY23} but realise it -- instead of the synchronous MPST framework -- in the asynchronous CFSM setting which needs completely different design and proof techniques. At the same time we go beyond the binary composition of asynchronous CFSMs of~\cite{BdLH19} and study multicomposition of CFSM systems. Clearly this goes also beyond the aforementioned papers~\cite{BLT20,BLP22b,BLT23} dealing with binary composition of synchronous CFSMs. In particular, in the asynchronous case different communication properties, like unspecified-reception freeness, are relevant. 
%In the present paper we investigate safety of PaI multicomposition for the
%asynchronous formalism of CFSMs going beyond the binary case studied in~\cite{BdLH19}.
%We reuse the PaI multicomposition \bfr idea introduced in~\cite{BDGY23}
%for a MPTS synchronous setting, \efr 
%but apply it to the \bfr asynchronous \efr framework of CFSMs 
%which  needs completely different design and proof techniques.
%and
%leads to significantly new challenges.  In particular,
%in the asynchronous case different communication properties,
%like freeness of unspecified receptions, appear. 

A crucial role in our approach to multicomposition is played by
\emph{connection policies} which can be individually chosen by the system designer on the basis of a given  concrete  \emph{connection model}.
A connection model describes architectural aspects of compositions.
It specifies which forwarding links between interface roles of different systems are meaningful from a static perspective.
The concrete behavioural instantiation of such links, in terms of
which message of an interface role, say $\HH$, is forwarded in which state of $\HH$ to which interface role of another system, is determined by a
connection policy which therefore also determines the construction of gateway CFSMs.
The \emph{multicomposition} of $n$ systems of CFSMs is
then simply defined by taking all CFSMs of the single systems
but replacing each CFSM of an interface participant by its gateway CFSM.
The use of connection models %describing static, architectural aspects of systems composition
is methodologically important since it is more likely that a connection policy complying with a connection model will satisfy desired communication properties.
For the proofs of our safety results, however, only the specifics of the chosen connection policy is relevant.
% Otherwise connection models are not relevant to our proofs.
%
%In other words, we assume the existence of some connection model with which the connection policy used for the multicomposition is compliant. However, the specifics of the connection model are irrelevant for the safety results.

We show that a number of relevant communication properties
(deadlock-freeness, orphan-message freeness, unspecified-reception freeness, and progress) are preserved by PaI multicomposition of CFSM systems
%via gateways 
whenever the particular property is satisfied also by the connection policy used, which is formalised as a CFSM system itself. 
Apart from orphan-message-freeness preservation we need, however, an additional assumption which requires that interface participants do not
have a state with at least one outgoing output action and one
outgoing input action, 
a condition referred to in the literature as {\em no-mixed-state}~\cite{CF05}.
 We shall provide counterexamples illustrating the role played by the no-mixed-state condition
in guaranteeing safety of composition.
 In contrast with deadlock-freeness, the stronger property of lock-freeness
will be shown (by means of a counterexample) not to be preserved in general, even in absence of of mixed-states.

%A counterexample will also be provided showing that,   even in the presence of no-mixed-states, the property of lock-freeness is, in general,
%not preserved. 

\medskip
\noindent
{\em Outline.} 
 The main ideas underlying PaI multicomposition are intuitively described
in \cref{sec:pai-multicomp}.  
In \cref{sect:cfsm} we recall  the definitions of communicating finite state machine, communicating system and their related notions.  
There we also provide the definitions of a number of relevant communication properties. 
In \cref{sec:opensys}, PaI multicomposition is   formally defined  on the basis  of the definitions
of connection policy and gateway. 
%\brc I would omit the next sentence since we have many examples. \erc
%An example is used throughout the paper to illustrate the various concepts introduced. 
Our main  results   are presented in~\cref{sec:preservation}
including counterexamples spotting the role of the no-mixed-state condition and a counterexample
for lock-freeness preservation. 
\cref{sect:conclusions} concludes with a brief summary, by pointing out a few more  
approaches to system composition, and with hints for future work. \\

\section{The PaI Approach to Multicomposition}
\label{sec:pai-multicomp}

%It is worth remarking that PaI applies to implemented closed systems
%(so enabling to look at them as virtually open) but,  at the same time, it can
%be considered also as a system-independent approach to the design of   
%open systems (simply requiring   the behaviour of the ``interface  participants'' 
%to be left unimplemented). 

\begin{wrapfigure}{r}{0.45\textwidth}
%\begin{figure}
    \vspace{-10mm}
    $
    \begin{array}{@{\hspace{0mm}}c@{\hspace{-2mm}}}
    \begin{array}{c@{\hspace{-2mm}}c}
    \text{\large $S_1$}
    &
 \begin{tikzpicture}[node distance=1.5cm,scale=1]
        \node (square-h) [draw,minimum width=0.8cm,minimum height=0.8cm] {\large $\hh$};
        \node [state] (h-a) [above of = square-h, draw=none] {};
        \node [state] (h-c) [left of = square-h, draw=none] {};
        \draw [-stealth] (h-a) --  node {$\msg[a]$} (square-h);
        \draw [-stealth] (square-h) --  node {$\msg[c]$} (h-c);
 \end{tikzpicture}
 \end{array}
 \hspace{4mm}
\begin{array}{c}
 \\
 \\
| \\
| \\
|\\
|\\
\end{array}
 \hspace{4mm}
 \begin{array}{c@{\hspace{-2mm}}c}
\begin{tikzpicture}[node distance=1.5cm,scale=1]
        \node (square-k) [draw,minimum width=0.8cm,minimum height=0.8cm] {\large $\kk$};
        \node [state] (k-b) [above of = square-k, draw=none] {};
        \node [state] (k-a) [right of = square-k, draw=none] {};
        \draw [-stealth] (k-b) --  node {$\msg[b]$} (square-k);
        \draw [-stealth] (square-k) --  node {$\msg[a]$} (k-a);
 \end{tikzpicture}
 &
 \text{\large $S_2$} 
 \end{array}
 \\[12mm]
\hspace{3mm}- - - -    \hspace{8mm}- - - - -  \\[-5mm]
\begin{array}{@{\hspace{0mm}}c@{\hspace{0mm}}c}
\\[4mm]
\text{\large $S_3\hspace{-2mm}$} 
&
 \begin{tikzpicture}[node distance=1.5cm,scale=1]
        \node (square-v) [draw,minimum width=0.8cm,minimum height=0.8cm] {\large $\ttv$};
        \node [state] (v-b) [left of = square-v, draw=none] {};
        \node [state] (v-a) [below of = square-v, draw=none] {};
        \draw [-stealth] (v-a) --  node {$\msg[a]$} (square-v);
        \draw [-stealth] (square-v) --  node {$\msg[b]$} (v-b);
 \end{tikzpicture}
 \end{array}
 \hspace{4mm}
\begin{array}{c}
 \\[-8mm]
| \\
| \\
| \\
|
\end{array}
 \hspace{4mm}
 \begin{array}{c@{\hspace{-2mm}}}
 \\[-6mm]
 \begin{tikzpicture}[node distance=1.5cm,scale=1]
        \node (square-w) [draw,minimum width=0.8cm,minimum height=0.8cm] {\large $\ttw$};
        \node [state] (w-c) [right of = square-w, draw=none] {};
        \node [state] (w-a) [below of = square-w, draw=none] {};
        \node [state] (w-b) [above right of = square-w, draw=none] {};
        \draw [-stealth] (w-c) --  node {$\msg[c]$} (square-w);
        \draw [-stealth] (square-w) --  node {$\msg[a]$} (w-a);
        \draw [stealth-] (square-w) --  node {$\msg[b]$} (w-b);
 \end{tikzpicture}
 \hspace{-6mm}
 \begin{array}{l}
 \\[6mm]
 \text{\large\ \ $S_4$}
 \end{array} 
 \end{array}
 \\[-6mm]
 \end{array}
 $
\caption{\label{fig:four-ips}
Four interface participants}
% \end{figure}
 \vspace{-5mm}
 \end{wrapfigure}
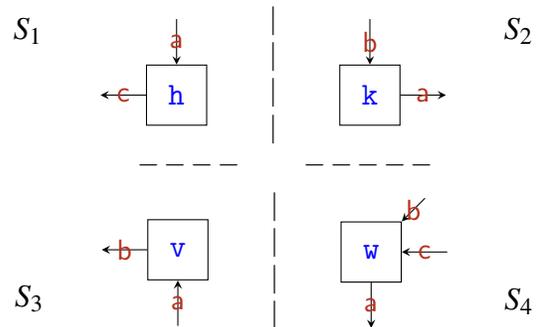

In order to illustrate the idea underlying \emph{PaI
  multicomposition}\footnote{It is of course possible to compose, two by two, several systems using binary composition,
but in that way -- by looking at systems as vertices and gateway connections as undirected edges  -- we can get only tree-like structures of systems.}, 
 we consider an example of~\cite{BDGY23} with
four systems $S_1$,  $S_2$, $S_3$ and $S_4$.
As shown in~\cref{fig:four-ips}, we have selected for each system one participant
as an interface, named $\hh$, $\kk$, $\ttv$ and $\ttw$.
As in~\cref{fig:bincomp}, we consider here only static aspects abstracting from dynamic issues, like   
the logical order of the exchanged messages,
 whose representation depends on the chosen formalism.

Following the PaI approach, the composition of the four
 systems above consists in replacing the participants  $\hh$, $\kk$, $\ttv$ and $\ttw$, chosen  as
interfaces, by gateways.  Note that a message, like $\msg[a]$ in $S_1$ sent to $\hh$, could be forwarded (unlike the binary case) to  different other gateways.
%gateways of different  sessions
This means that a \emph{connection policy} has to be set up
in order to  appropriately define the gateways. Such a policy primarily depends on
which partner is chosen for the current message to be exchanged.  

\begin{figure}[ht]%{c}{0.6\textwidth}
    \centering{
    $\begin{array}{c@{\qquad\qquad\qquad\qquad\qquad}c}
 \begin{tikzpicture}[node distance=1.5cm,scale=1]
        \node (square-h) [draw,minimum width=0.8cm,minimum height=0.8cm] {\large $\hh$};
%        \node [state] (h-a) [above of = square-h, draw=none] {};
%        \node [state] (h-c) [left of = square-h, draw=none] {};
%        \draw [-stealth] (h-a) --  node {$a$} (square-h);
%        \draw [-stealth] (square-h) --  node {$c$} (h-c);
        \node (square-k) [draw,minimum width=0.8cm,minimum height=0.8cm, right of = square-h] {\large $\kk$};
%        \node [state] (k-b) [above of = square-k, draw=none] {};
%        \node [state] (k-a) [right of = square-k, draw=none] {};
%        \draw [-stealth] (k-b) --  node {$\msg[b]$} (square-k);
%        \draw [-stealth] (square-k) --  node {$a$} (k-a);
        \node (square-v)  [draw,minimum width=0.8cm,minimum height=0.8cm, below of = square-h, yshift=-4mm] {\large $\ttv$};
%        \node [state] (v-b) [left of = square-v, draw=none] {};
%        \node [state] (v-a) [below of = square-v, draw=none] {};
%        \draw [-stealth] (v-a) --  node {$a$} (square-v);
%        \draw [-stealth] (square-v) --  node {$b$} (v-b);
        \node (square-w)  [draw,minimum width=0.8cm,minimum height=0.8cm, below of = square-k, yshift=-4mm] {\large $\ttw$};
%        \node [state] (w-c) [right of = square-w, draw=none] {};
%        \node [state] (w-a) [below of = square-w, draw=none] {};
%        \node [state] (w-b) [above right of = square-w, draw=none] {};
%        \draw [-stealth] (w-c) --  node {$c$} (square-w);
%        \draw [-stealth] (square-w) --  node {$a$} (w-a);
%        \draw [stealth-] (square-w) --  node {$b$} (w-b);
        %
      %  \draw (square-h)  to[out=-90,in=90]   node {} (square-w);
       % \draw (square-k) to[out=-90,in=90]  node {} (square-v);
      %  \draw (square-h) to[out=0,in=180]  node {} (square-w);
        \draw[-stealth]  (square-w) to[out=-135,in=-45]  node {$\msg[b]$} (square-v);
      %  \draw (square-k) to[out=-135,in=45]  node {} (square-v);
        %
%        \draw (0,0.4)[dotted,thick]  --  (0.4,0); % carrying a inside h  
%        \draw (-0.4,0)[dotted,thick]  --  (0,-0.4); % carrying c inside h
%        %
%        \draw (1.5,0.4)[dotted,thick]  --  (1.1,0); % carrying b inside k
%        \draw (1.5,-0.4)[dotted,thick]  --  (1.9,0); % carrying a inside k
%        %
%        \draw (0.4,-1.9)[dotted,thick]  --  (0,-2.3); % carrying a inside v
%        \draw (0,-1.5)[dotted,thick]  --  (-0.4,-1.9); % carrying k's b inside v
%        \draw (0.4,-2.3)[dotted,thick]  --  (-0.4,-1.9); % carrying b inside v
%        %
%        \draw (1.5,-1.5)[dotted,thick]  --  (1.5,-2.3); % carrying a inside w
%        \draw (1.1,-1.9)[dotted,thick]  --  (1.9,-1.9); % carrying c inside w
%        \draw (1.1,-2.3)[dotted,thick]  to[out=30,in=260]  (1.9,-1.5); % carrying b inside w
        %
        %
        \draw [-stealth] (0.4,0)  -- node {$\msg[a]$}  (1.5,-1.5 ); % carryng a from h to w    
         \draw [stealth-] (0,-0.4)  -- node {$\msg[c]$} (1.1,-1.9); % carryng c from w to h 
        \draw [-stealth] (0.4,-1.9)  --  node {$\msg[a]$} (1.5,-0.4 ); % carrying a from v to k
        \draw [stealth-] (0,-1.5)  --  node {$\msg[b]$} (1.1,0); % carrying  b from k to v
 \end{tikzpicture}
&
 \begin{tikzpicture}[node distance=1.5cm,scale=1]
        \node (square-h) [draw,minimum width=0.8cm,minimum height=0.8cm] {\large $\HH$};
        %\node [state] (h-a) [above of = square-h, draw=none] {};
        %\node [state] (h-c) [left of = square-h, draw=none] {};
        %\draw [-stealth] (h-a) --  node {$a$} (square-h);
        %\draw [-stealth] (square-h) --  node {$c$} (h-c);
        \node (square-k) [draw,minimum width=0.8cm,minimum height=0.8cm, right of = square-h] {{\large $\KK$}};
%        \node [state] (k-b) [above of = square-k, draw=none] {};
%        \node [state] (k-a) [right of = square-k, draw=none] {};
%        \draw [-stealth] (k-b) --  node {$b$} (square-k);
%        \draw [-stealth] (square-k) --  node {$a$} (k-a);
        \node (square-v)  [draw,minimum width=0.8cm,minimum height=0.8cm, below of = square-h, yshift=-4mm] {{\large $\ttv$}};
%        \node [state] (v-b) [left of = square-v, draw=none] {};
%        \node [state] (v-a) [below of = square-v, draw=none] {};
%        \draw [-stealth] (v-a) --  node {$a$} (square-v);
%        \draw [-stealth] (square-v) --  node {$b$} (v-b);
        \node (square-w)  [draw,minimum width=0.8cm,minimum height=0.8cm, below of = square-k, yshift=-4mm] {{\large $\ttw$}};
%        \node [state] (w-c) [right of = square-w, draw=none] {};
%        \node [state] (w-a) [below of = square-w, draw=none] {};
%        \node [state] (w-b) [above right of = square-w, draw=none] {};
%        \draw [-stealth] (w-c) --  node {$c$} (square-w);
%        \draw [-stealth] (square-w) --  node {$a$} (w-a);
%        \draw [stealth-] (square-w) --  node {$b$} (w-b);
        %
        \draw [-stealth] (square-w) to[out=-135,in=-45]  node {$\msg[b]$} (square-v);
        %
%        \draw (0,0.4)[dotted,thick]  --  (0.4,0); % carrying a inside h  
%        \draw (-0.4,0)[dotted,thick]  --  (0,-0.4); % carrying c inside h
%        %
%        \draw (1.5,0.4)[dotted,thick]  --  (1.1,0); % carrying b inside k
%        \draw (1.5,-0.4)[dotted,thick]  --  (1.9,0); % carrying a inside k
%        %
%        \draw (0.4,-1.9)[dotted,thick]  --  (0,-2.3); % carrying a inside v
%        \draw (0,-1.5)[dotted,thick]  --  (-0.4,-1.9); % carrying k's b inside v
%        \draw (0.4,-2.3)[dotted,thick]  --  (-0.4,-1.9); % carrying b inside v
%        %
%        \draw (1.5,-1.5)[dotted,thick]  --  (1.5,-2.3); % carrying a inside w
%        \draw (1.1,-1.9)[dotted,thick]  --  (1.9,-1.9); % carrying c inside w
%        \draw (1.1,-2.3)[dotted,thick]  to[out=30,in=260]  (1.9,-1.5); % carrying b inside w
        %
        %
        \draw [-stealth] (0.4,-1.9) to[out=45,in=90] node {$\msg[a]$} (1.5,-1.5 ); % carryng a from v to w    
         \draw [stealth-] (0,-0.4)  -- node {$\msg[c]$} (1.1,-1.9) ; % carryng c from w to h 
        \draw [-stealth]  (0.4,0)  to[out=-45,in=-90]  node {$\msg[a]$} (1.5,-0.4 ); % carrying a from h to k
        \draw [stealth-] (0,-1.5)  --  node {$\msg[b]$} (1.1,0); % carrying  b from k to v
 \end{tikzpicture}\\
 \begin{array}{c}
  \\[1mm]
 (\text{Choice A})
 \end{array}&
 \begin{array}{c}
 \\[1mm]
 (\text{Choice B})
 \end{array}
 \end{array}
 $
 }
 \caption{\label{fig:choicesAB} Two possible choices of partners.}\label{fig:twocm}
 \end{figure}
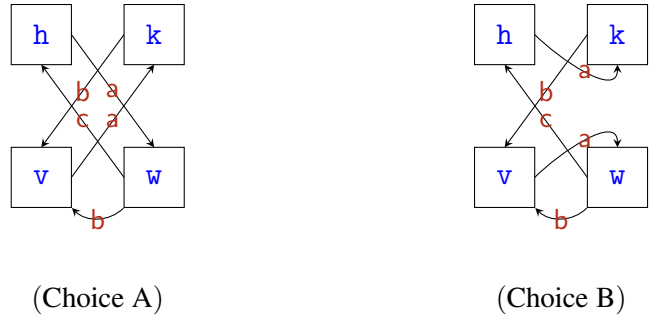

 For what concerns  the present example,
 one could decide that message $\msg[a]$ received by $\hh$ has to be forwarded
 to $\ttw$; the $\msg[a]$ received by $\ttv$  to $\kk$; the $\msg[b]$ received by $\kk$  and $\ttw$ to $\ttv$; 
 the $\msg[c]$ received by $\ttw$ to $\hh$.
 Another possible choice %(see policy B  in Figure~\ref{fig:twocm})) 
could be similar to the previous one but for the forwarding
of the messages $\msg[a]$: the one received by $\HH$ could be forwarded now to $\KK$
whereas the one received by $\ttv$ could be forwarded to $\ttw$. 
 Such different ``choices of partners'', that we formalise %subsequently
 by introducing the notion of {\em connection model}, can be graphically represented, respectively, by  Choice A and Choice B in
\cref{fig:choicesAB}.
%For the present example several other connection models would be possible,
%for instance the one allowing to forward the $\msg[a]$ from $\hh$
%either to $\kk$ or to $\ttw$.
%Since a connection model abstracts from the dynamic of participants, a connection policy does not depend in general only on the connection model taken into account.

%%% CONNECTION MODELS
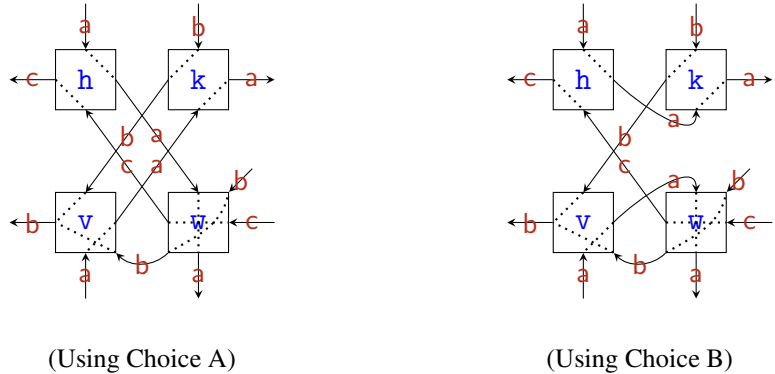
\begin{figure}[t]%{c}{0.6\textwidth}
\vspace{-12mm}
    \centering{
    $\begin{array}{c@{\quad\qquad}c}
 \begin{tikzpicture}[node distance=1.5cm,scale=1]
        \node (square-h) [draw,minimum width=0.8cm,minimum height=0.8cm] {\large $\hh$};
        \node [state] (h-a) [above of = square-h, draw=none] {};
        \node [state] (h-c) [left of = square-h, draw=none] {};
        \draw [-stealth] (h-a) --  node {$\msg[a]$} (square-h);
        \draw [-stealth] (square-h) --  node {$\msg[c]$} (h-c);
        \node (square-k) [draw,minimum width=0.8cm,minimum height=0.8cm, right of = square-h] {\large $\kk$};
        \node [state] (k-b) [above of = square-k, draw=none] {};
        \node [state] (k-a) [right of = square-k, draw=none] {};
        \draw [-stealth] (k-b) --  node {$\msg[b]$} (square-k);
        \draw [-stealth] (square-k) --  node {$\msg[a]$} (k-a);
        \node (square-v)  [draw,minimum width=0.8cm,minimum height=0.8cm, below of = square-h, yshift=-4mm] {\large $\ttv$};
        \node [state] (v-b) [left of = square-v, draw=none] {};
        \node [state] (v-a) [below of = square-v, draw=none] {};
        \draw [-stealth] (v-a) --  node {$\msg[a]$} (square-v);
        \draw [-stealth] (square-v) --  node {$\msg[b]$} (v-b);
        \node (square-w)  [draw,minimum width=0.8cm,minimum height=0.8cm, below of = square-k, yshift=-4mm] {\large $\ttw$};
        \node [state] (w-c) [right of = square-w, draw=none] {};
        \node [state] (w-a) [below of = square-w, draw=none] {};
        \node [state] (w-b) [above right of = square-w, draw=none] {};
        \draw [-stealth] (w-c) --  node {$\msg[c]$} (square-w);
        \draw [-stealth] (square-w) --  node {$\msg[a]$} (w-a);
        \draw [stealth-] (square-w) --  node {$\msg[b]$} (w-b);
        %
      %  \draw (square-h)  to[out=-90,in=90]   node {} (square-w);
       % \draw (square-k) to[out=-90,in=90]  node {} (square-v);
      %  \draw (square-h) to[out=0,in=180]  node {} (square-w);
        \draw [-stealth] (square-w) to[out=-135,in=-45]  node {$\msg[b]$} (square-v);
      %  \draw (square-k) to[out=-135,in=45]  node {} (square-v);
        %
        \draw (0,0.4)[dotted,thick]  --  (0.4,0); % carrying a inside h  
        \draw (-0.4,0)[dotted,thick]  --  (0,-0.4); % carrying c inside h
        \draw (1.5,0.4)[dotted,thick]  --  (1.1,0); % carrying b inside k
        \draw (1.5,-0.4)[dotted,thick]  --  (1.9,0); % carrying a inside k
        \draw (0.4,-1.9)[dotted,thick]  --  (0,-2.3); % carrying a inside v
        \draw (0,-1.5)[dotted,thick]  --  (-0.4,-1.9); % carrying k's b inside v
        \draw (0.4,-2.3)[dotted,thick]  --  (-0.4,-1.9); % carrying b inside v
        \draw (1.5,-1.5)[dotted,thick]  --  (1.5,-2.3); % carrying a inside w
        \draw (1.1,-1.9)[dotted,thick]  --  (1.9,-1.9); % carrying c inside w
        \draw (1.1,-2.3)[dotted,thick]  to[out=30,in=260]  (1.9,-1.5); % carrying b inside w
        \draw[-stealth]   (0.4,0)  -- node  {$\msg[a]$}   (1.5,-1.5 ); % carryng a from h to w    
         \draw  [stealth-]  (0,-0.4)  --  node  {$\msg[c]$}  (1.1,-1.9); % carryng c from w to h 
        \draw [-stealth]  (0.4,-1.9)  --   node  {$\msg[a]$} (1.5,-0.4 ); % carrying a from v to k
        \draw [stealth-]  (0,-1.5)  --  node {$\msg[b]$} (1.1,0); % carrying  b from k to v
 \end{tikzpicture}
& 
\begin{tikzpicture}[node distance=1.5cm,scale=1]
        \node (square-h) [draw,minimum width=0.8cm,minimum height=0.8cm] {\large $\HH$};
        \node [state] (h-a) [above of = square-h, draw=none] {};
        \node [state] (h-c) [left of = square-h, draw=none] {};
        \draw [-stealth] (h-a) --  node {$\msg[a]$} (square-h);
        \draw [-stealth] (square-h) --  node {$\msg[c]$} (h-c);
        \node (square-k) [draw,minimum width=0.8cm,minimum height=0.8cm, right of = square-h] {\large $\KK$};
        \node [state] (k-b) [above of = square-k, draw=none] {};
        \node [state] (k-a) [right of = square-k, draw=none] {};
        \draw [-stealth] (k-b) --  node {$\msg[b]$} (square-k);
        \draw [-stealth] (square-k) --  node {$\msg[a]$} (k-a);
        \node (square-v)  [draw,minimum width=0.8cm,minimum height=0.8cm, below of = square-h, yshift=-4mm] {\large $\ttv$};
        \node [state] (v-b) [left of = square-v, draw=none] {};
        \node [state] (v-a) [below of = square-v, draw=none] {};
        \draw [-stealth] (v-a) --  node {$\msg[a]$} (square-v);
        \draw [-stealth] (square-v) --  node {$\msg[b]$} (v-b);
        \node (square-w)  [draw,minimum width=0.8cm,minimum height=0.8cm, below of = square-k, yshift=-4mm] {\large $\ttw$};
        \node [state] (w-c) [right of = square-w, draw=none] {};
        \node [state] (w-a) [below of = square-w, draw=none] {};
        \node [state] (w-b) [above right of = square-w, draw=none] {};
        \draw [-stealth] (w-c) --  node {$\msg[c]$} (square-w);
        \draw [-stealth] (square-w) --  node {$\msg[a]$} (w-a);
        \draw [stealth-] (square-w) --  node {$\msg[b]$} (w-b);
        \draw [-stealth] (square-w) to[out=-135,in=-45]  node {$\msg[b]$} (square-v);
        \draw (0,0.4)[dotted,thick]  --  (0.4,0); % carrying a inside h  
        \draw (-0.4,0)[dotted,thick]  --  (0,-0.4); % carrying c inside h
        \draw (1.5,0.4)[dotted,thick]  --  (1.1,0); % carrying b inside k
        \draw (1.5,-0.4)[dotted,thick]  --  (1.9,0); % carrying a inside k
        \draw (0.4,-1.9)[dotted,thick]  --  (0,-2.3); % carrying a inside v
        \draw (0,-1.5)[dotted,thick]  --  (-0.4,-1.9); % carrying k's b inside v
        \draw (0.4,-2.3)[dotted,thick]  --  (-0.4,-1.9); % carrying b inside v
        \draw (1.5,-1.5)[dotted,thick]  --  (1.5,-2.3); % carrying a inside w
        \draw (1.1,-1.9)[dotted,thick]  --  (1.9,-1.9); % carrying c inside w
        \draw (1.1,-2.3)[dotted,thick]  to[out=30,in=260]  (1.9,-1.5); % carrying b inside w
        \draw [-stealth]  (0.4,-1.9) to[out=45,in=90]  node [pos=0.6] {$\msg[a]$} (1.5,-1.5 ); % carryng a from v to w    
         \draw [stealth-]  (0,-0.4)  --  node  {$\msg[c]$} (1.1,-1.9); % carryng c from w to h 
        \draw [-stealth]  (0.4,0)  to[out=-45,in=-90]  node [pos=0.6]  {$\msg[a]$} (1.5,-0.4 ); % carrying a from h to k
        \draw [stealth-]  (0,-1.5)  -- node  {$\msg[b]$}  (1.1,0); % carrying  b from k to v
 \end{tikzpicture}\\[-4mm]
 \text{\small (Using Choice A)}
 &
 \text{\small (Using Choice B)}
 \end{array}$
  }
  \vspace{-1mm}
  \caption{Two possible PaI multicompositions via gateways}
  \label{fig:multiconnection}
 \end{figure}

 The  architecture  of the resulting composed systems, according to the particular choices of partners  (i.e.\ connection models), are 
represented  by the diagrams  in~\cref{fig:multiconnection}.

 In both drawings of~\cref{fig:multiconnection}, the names $\hh, \kk$, etc.\  do now represent gateways.
It is important to see that even if the original CFSMs for the participants
in the single systems, like the CFSM for $\ttv$ in $S_3$, are given, the connection models and the drawings in
~\cref{fig:multiconnection}
do not always determine how 
%provide  what 
a gateway CFSM,  modelling the dynamic forwarding strategy,  should look like. 
%enough information on how a gateway CFSM, 
% modelling the dynamic forwarding strategy,  should look like.
This can be illustrated by looking at message $\msg[b]$ and participant $\ttv$.
No matter whether we consider Choice A or Choice B it is
not determined when the gateway for $\ttv$ will accept $\msg[b]$ 
from $\kk$ and when from $\ttw$.
For instance,  
a message $\msg[b]$ from $\ttw$ could be accepted by $\ttv$ only after two $\msg[b]$'s are received from $\kk$.
Therefore, a given choice of partners needs, in general, to be ``refined'' -- according to the formalism taken into account -- into a specific 
{\em connection policy}  taking care of the dynamic choice of partners.

This PaI approach to multicomposition has been exploited in~\cite{BDGY23}
for a MPST formalism with synchronous communications.
 We are now going to realise PaI multicomposition
in the context of CFSM systems with asynchronous communications.

%!TEX root = Main-asynchCFSM-multicomp.tex

\section{Systems of Communicating Finite State Machines}
\label{sect:cfsm}

Communicating Finite State Machines (CFSMs) is a widely investigated
formalism for the description and analysis of distributed systems, originally proposed in \cite{BZ83}.
CFSMs are a variant of finite state I/O-automata that represent processes which communicate by asynchronous exchanges of messages via FIFO channels. 
We now recall (partly following \cite{CF05,DY12,TY15,BdLH19}) the definitions of CFSM and  system of CFSMs.

We assume given a countably infinite set  
$\roles_\mathfrak{U}$ of participant names (ranged over by $\ttp,\ttq,\ttr,\HH,\KK,\ldots$) and a countably infinite alphabet $\mathbb{A}_\mathfrak{U}$ 
of messages (ranged over by $\msg[a]$, $\msg[b]$, $\msg[c]$, $\msg[l]$, $\msg[m],\ldots$).\\

\begin{definition}[CFSM]\label{def:cfsm}%\hfill\\
Let $\roles$  and $\mathbb{A}$ be finite subsets of $\roles_\mathfrak{U}$ and $\mathbb{A}_\mathfrak{U}$ respectively.
\begin{enumerate}[i)] 
\item
The set $C_\roles$ of {\em channels} over $\roles$ is defined by\ \
$C_\roles=\Set{\ttp\ttq \mid \ttp,\ttq\in \roles, \ttp\neq\ttq}$
\item
The set $\mathit{Act}_{\roles,\mathbb{A}}$ of {\em actions}  over $\roles$ and $\mathbb{A}$ is defined by\ \
$\textit{Act}_{\roles,\mathbb{A}} = C_\roles\times\Set{!,?}\times\mathbb{A}$

The {\em subject} of an output action $\ttp\ttq!\msg[m]$ and of an input action $\ttq\ttp?\msg[m]$ is 
$\ttp$.
\item
\label{def:cfsm-iii}
A {\em communicating finite-state machine over} $\roles$ \emph{and} $\mathbb{A}$
is a finite transition system given by a tuple\\
\centerline{ $M=(Q,q_0,\mathbb{A},\delta)$ }
where $Q$ is a finite set of states, $q_0\in Q$ is the initial state, and
$\delta\subseteq Q\times\textit{Act}_{\roles,\mathbb{A}}\times Q$ is a set of transitions
such that all the actions have the same subject, to which we refer as the {\em name} of $M$.
\end{enumerate}
\end{definition}
\noindent
We shall write $M_{\ttp}$ to denote a CFSM with name $\ttp$. 
Where no ambiguity arises we shall refer to a CFSM by its name.

Notice that the above definition of CFSM is generic with respect to the underlying sets
$\roles$ and $\mathbb{A}$.
This is necessary,  since we shall not deal with a single system of CFSMs but with an arbitrary number of  systems of CFSMs that can be {\em composed}.
We shall write $C$ and $\mathit{Act}$ instead of $C_\roles$ and $\mathit{Act}_{\roles,\mathbb{A}}$ when no ambiguity can arise.
%\brc Let us check whether we ever use $C$ and $\mathit{Act}$.
%But even then, I would prefer to see $C_\roles$ and $\mathit{Act}_{\roles,\mathbb{A}}$ which is minimal longer.\erc
 We assume $\elle,\elle',\ldots$ to range over $\textit{Act}$;
$\varphi,\varphi',\ldots$ to range over $\textit{Act}^*$ (the set of finite words over $\textit{Act}$), and
$w,w',\ldots$ to range over $\mathbb{A}^*$ (the set of finite words over $\mathbb{A}$).
The symbol $\varepsilon\,(\notin \mathbb{A}\cup\textit{Act})$ denotes the empty word and $\mid v\mid$ the lenght of a word $v\in \textit{Act}^*\cup\mathbb{A}^*$.

%%>>>>>>>> DEFINITIONS NOT USED in the present paper
%Given a word $v$ with prefix $v'$, i.e. such that $v=v'\cdot v''$ for a certain $v''$, we define $v\setminus v' =v''$.
%Moreover, given a  word $v$ with $\msg[a]$ as last  element, i.e. $v=v'\cdot \msg[a]$ for a certain (possibly empty) $v'$, we define
%$\mathsf{init}(v) = v'$ and  $\mathsf{last}(v) = \msg[a]$. 
%Moreover, we shall denote by $\widetilde{v}$ the reverse of the word $v$. \\

The transitions of a CFSM are labelled by actions; a label $\tts\ttr!\msg[a]$ represents
the asynchronous sending of message $\msg[a]$ from machine $\tts$ to $\ttr$ through channel $\tts\ttr$ and, dually,
$\tts\ttr?\msg[a]$ represents the reception (consumption) of $\msg[a]$ by $\ttr$ from channel
$\tts\ttr$. 

 Given a CFSM $M=(Q,q_0,\mathbb{A},\delta)$,
we also define \\
\centerline{$\inn{M}=\Set{\msg[a] \mid (\_,\_\,\_?\msg[a],\_)\in \delta }$
\quad \text{ and }\quad $\outt{M}=\Set{\msg[a] \mid (\_,\_\,\_!\msg[a],\_)\in \delta }$.}
 If $M$ is a CFSM with name $\ttp$, we also write $\inn{\ttp}$ for $\inn{M}$ and
 $\outt{\ttp}$ for $\outt{M}$.
  Note that, in concrete examples, the name of a CFSM together with its input and output messages can be graphically depicted as in~\cref{fig:four-ips}.

%\brc
%(i) I wonder whether we need somewhere $\inn{M} \cap \outt{M} = \emptyset$? Probably not. 
%(ii) I use the notation $\inn{p}$, $\outt{p}$ in~\cref{ex:scm}.
%And if you like it you can use it at many places. \bfr OK\efr
%\erc

%We write $\lang{M}\subseteq\textit{Act}^*$ for
%the language over $\textit{Act}$ accepted by the automaton corresponding
%to machine $M$, where each state of $M$ is an accepting state. 
A state
$q\in Q$ with no outgoing transition is {\em final}; 
$q$ is a {\em sending} (resp. {\em receiving}) state if it is not final and
all outgoing transitions are labelled with sending (resp. receiving) actions;
$q$  is a {\em mixed} state if there are at least two outgoing transitions such that one is labelled with a sending action and the other one is labelled with a receiving action.

%
%  ?!-DETERMINISM DEFINITION <<<<<<<<<<<<<<<<<<<<<<<<
% 
%\vspace{2mm}
%A CFSM $M = (Q,q_0,\mathbb{A},\delta)$ is:
%\begin{enumerate}[a)]
%\item
% {\em deterministic} if for all transitions:\quad % states $q\in Q$ and all actions $\elle$: 
%$(q,\elle, q'), (q,\elle,q'')\in \delta$ imply $q'=q''$;
%\item
%{\em ?-deterministic} (resp. {\em !-deterministic}) if for all transitions:\\ % all states  $q\in Q$ and all actions:\\
%$\qquad$ $(q,\ttr\tts?\msg[a], q'), (q,\ttp\ttq?\msg[a],q'')\in \delta$ (resp. $(q,\ttr\tts!\msg[a], q'), (q,\ttp\ttq!\msg[a],q'')\in \delta$) imply $q'=q''$;\footnote{Note that, by Definition \ref{def:cfsm}(\ref{def:cfsm-iii}), we have
%necessarily that $\tts=\ttq$ in the clause for ?-determinism and $\ttr=\ttp$ in the one for
%!-determinism.}
%\item
%{\em ?!-deterministic} if it is both ?-deterministic and !-deterministic.
%\end{enumerate}
%
%The notion of ?!-deterministic machine is more demanding than in usual CFSM settings. It will be needed in order to guarantee preservation of communication properties when systems are connected. 
%Note that a ?!-deterministic CFSM is also deterministic, but the converse does not hold
%(since the channel names are abstracted away in the definition of ?!-determinism). \\

A {\em communicating system}, called ``protocol'' in \cite{BZ83}, is a finite set of CFSMs.
% over some vocabulary of messages such that senders and receivers are identified by the 
%names of CFSMs. 
 In~\cite{CF05,DY12,TY15} the names of the CFSMs in a system are called {\em roles}. In the present paper we call them {\em participants}.

The dynamics of a system is formalised as a transition relation on configurations, where a configuration is a
pair of tuples: a tuple of states of the machines in the system and a tuple of buffers representing the content of the channels. 

\begin{definition}[Communicating system and configuration]%\hfill\\
Let $\roles$  and $\mathbb{A}$ be as in Def.~\ref{def:cfsm}.
\begin{enumerate}[i)]
\item
A {\em communicating system (CS)
over} $\roles$ \emph{and} $\mathbb{A}$ is a  set  %tuple 
$S= (M_\ttp)_{\ttp\in\roles}$
%\centerline{$S= (M_\ttp)_{\ttp\in\roles}$}
where\\
%\\
%-  $\roles\subseteq_{\text{fin}}\roles_\mathfrak{U}$   is the set of {\em roles} (participants) of $S$, and\\
for each $\ttp\in \roles$,
$M_\ttp=(Q_\ttp,q_{0\ttp},\mathbb{A},\delta_\ttp)$ is a CFSM  over $\roles$ and $\mathbb{A}$.
\item
A {\em configuration} of a system $S$ is a pair $s = (\vec{q},\vec{w})$
%\centerline{$s = (\vec{q},\vec{w})$}
where\\
\centerline{$\vec{q}= (q_\ttp)_{\ttp\in\roles}$ with $q_\ttp \in Q_\ttp$,
\qquad and \qquad  $\vec{w}  = (w_{\ttp\ttq})_{\ttp\ttq\in C}$ with $w_{\ttp\ttq}\in\mathbb{A^*}$.}
%\begin{itemize}
%\item[-]  $\vec{q}= (q_\ttp)_{\ttp\in\roles}$ with $q_\ttp \in Q_\ttp$,
%\item[-]  $\vec{w}  = (w_{\ttp\ttq})_{\ttp\ttq\in C}$ with $w_{\ttp\ttq}\in\mathbb{A^*}$.
%\end{itemize}

The component $\vec{q}$ is the {\em control state\/} of the system and $q_\ttp \in Q_\ttp$ is the 
{\em local state\/} of machine $M_\ttp$. 
The component $\vec{w}$ represents the state of the channels of the system and $w_{\ttp\ttq} \in \mathbb{A}^*$ is the state of the channel $\ttp\ttq$, i.e. the messages sent from $\ttp$ to $\ttq$. The initial configuration of $S$ is $s_0=  (\vec{q_0},\vec{\varepsilon})$
with $\vec{{q_0}} = (q_{0_\ttp})_{\ttp\in\roles}$.
\end{enumerate}
\end{definition}

\noindent
In the following we shall often denote a communicating system $(M_{\ttp})_{\ttp\in \Set{\ttr_i}_{i\in I}}$ by $(M_{\ttr_i})_{i\in I}$.

\begin{definition}[Reachable configuration]% \hfill\\
Let $S$ be a communicating system over $\roles$ and $\mathbb{A}$, and let $s= (\vec{q},\vec{w})$ and $s'= (\vec{q'},\vec{w'})$ 
be two configurations of $S$. %\\
Configuration $s'$ {\em is reachable from} $s$
{\em by firing  a transition} with action $\elle$, written $s\lts{\elle}s'$, if there is $\msg[a]\in\mathbb{A}$
such that one of the following conditions holds:
%\begin{center}
%\begin{tabular}{ r l r l}
%$1.$ & $\elle = \tts\ttr!\msg[a]$ and $(q_\tts,\elle,q'_\tts)\in\delta_\tts$ and &  
%$2.$ & $\elle = \tts\ttr?\msg[a]$ and $(q_\ttr,\elle,q'_\ttr)\in\delta_\ttr$ and\\
%        & $a)$ for all $\ttp\neq\tts: ~ q'_\ttp =  q_\ttp$  and &
%        & $a)$ for all $\ttp\neq\ttr: ~ q'_\ttp =  q_\ttp$  and \\
%        & $b)$ $w'_{\tts\ttr} =  w_{\tts\ttr}\cdot \msg[a]$ and for all $\ttp\ttq\neq\tts\ttr: ~ w'_{\ttp\ttq} =  w_{\ttp\ttq}$; &
%        & $b)$  $w_{\tts\ttr} =  \msg[a]\cdot w'_{\tts\ttr}$ and for all $\ttp\ttq\neq\tts\ttr: ~w'_{\ttp\ttq} =  w_{\ttp\ttq}$.
%\end{tabular}
%\end{center}
\begin{enumerate}
\item
$\elle = \tts\ttr!\msg[a]$ and $(q_\tts,\elle,q'_\tts)\in\delta_\tts$ and
\qquad \begin{enumerate}[a)]
\item
for all $\ttp\neq\tts: ~ q'_\ttp =  q_\ttp$  and
\item
$w'_{\tts\ttr} =  w_{\tts\ttr}\cdot \msg[a]$ and for all $\ttp\ttq\neq\tts\ttr: ~ w'_{\ttp\ttq} =  w_{\ttp\ttq}$;
\end{enumerate}
\item 
$\elle = \tts\ttr?\msg[a]$ and $(q_\ttr,\elle,q'_\ttr)\in\delta_\ttr$ and
\begin{enumerate}[a)]
\item
for all $\ttp\neq\ttr: ~ q'_\ttp =  q_\ttp$  and
\item
$w_{\tts\ttr} =  \msg[a]\cdot w'_{\tts\ttr}$ and for all $\ttp\ttq\neq\tts\ttr: ~w'_{\ttp\ttq} =  w_{\ttp\ttq}$.
\end{enumerate}
\end{enumerate}
We write $s\lts{}s'$ if there exists $\elle$ such that  $s\lts{\elle}s'$
 and we write $s\notlts{}\hspace{2mm}$ if no $s'$ and no $\elle$ exist with
$s\lts{\elle}s'$.
As usual, we denote the reflexive and transitive 
closure of $\lts{}$ by $\to^*$.
The set of {\em reachable configurations} of S is $\RS(S) = \Set{s \mid s_0 \to^* s}.$
\end{definition}
\noindent
According to the above definition, communication happens via buffered channels following the FIFO principle.\\

The overall behaviour of a system can be described (at least) by the traces of configurations that are reachable from a distinguished initial one. Configurations may exhibit some pathological properties, like various forms of {\em deadlock} or {\em progress violation}, channels containing messages that will never be consumed ({\em orphan messages}) or just sent to a participant who is expecting another message to come ({\em unspecified receptions}). The goal of the analysis of
communicating systems is to check whether such kinds
of configurations are reachable or not. Although the desirable system properties are undecidable in general~\cite{BZ83}, sufficient conditions are known that are effectively checkable
relying, for instance, on half-duplex communication~\cite{CF05}, on the form of network topologies~\cite{DBLP:conf/concur/ClementeHS14}, or on synchronous compatibility checking~\cite{HB18}.

We formalise now a number of relevant communication properties for systems of CFSMs
that we shall deal with in the present paper.  

\begin{definition}[Communication properties]%\hfill\\
\label{def:safeness}
Let $S$ be a communicating system, and let $s= (\vec{q},\vec{w})$ be a configuration of $S$.
\begin{enumerate}[i)]
\item
\label{def:safeness-i}
$s$ is a {\em deadlock configuration} of $S$ if \hspace{2mm}
$\vec{w}=\vec{\varepsilon}\quad\text{and}\quad \forall \ttp\in\roles.~q_\ttp \text{ is a receiving state}$.\\
I.e. all buffers are empty, but all machines are waiting for a message.\\
We say that $S$ is {\em deadlock-free} whenever, for any $s\in \RS(S)$, $s$ is not a  deadlock configuration.

%\item
%\label{def:safeness-i}
%$s$ is a {\em deadlock configuration} if $s\, \not\!\!\lts{}$ and either
%\begin{enumerate}[a)]
%\item $\exists \ttr\in\roles$ such that $q_\ttr \lts{\ttr\tts?a} q'_\ttr$ , or
%\item
%\label{def:safeness-wnotem}
%$\vec{w}\neq\vec{\varepsilon}$ 
%\end{enumerate}
%i.e. $s$ is stuck  because all machines which are not in a final state are in a receiving state waiting for messages that cannot be read from the buffer; moreover if  $\vec{q}$ is final all buffers are empty.\\
%We say that $S$ is {\em deadlock-free} whenever, for any $s\in \RS(S)$, $s$ is not a  deadlock configuration.

\item
$s$ is an {\em  orphan-message  configuration} of $S$ if \hspace{2mm}
$\forall \ttp\in\roles. ~ q_\ttp \text{ is final} \quad\text{and}\quad  \vec{w}\neq \vec{\varepsilon}$.\\
I.e. each machine is in a final state, but there is still  at least one non-empty buffer.
We say that $S$ is {\em orphan-message free} whenever, for any $s\in \RS(S)$, $s$ is not an orphan-message configuration.

\item
\label{def:safeness-ur}
$s$ is an {\em unspecified reception configuration} of $S$  if ~$\exists \ttr \in\roles$ such that  
\begin{enumerate}[a)]
\item
%$\exists \ttr \in\roles. ~ 
$q_\ttr \text{ is a receiving state}$; and
\item
$\forall\tts\in\roles.[~(q_\ttr,\tts\ttr?\msg[a],q'_\ttr)\in\delta_\ttr  \implies
(|w_{\tts\ttr}| > 0~~\wedge~~ w_{\tts\ttr}\not\in  \msg[a]\cdot\mathbb{A}^*)~ ]$.
\end{enumerate}
I.e. there is a receiving  state $q_\ttr$ 
which is prevented from
receiving any message from any of its buffers.
(In other words, in each channel $\tts\ttr$ from which role $\ttr$ could consume there
is a message which cannot be received by $\ttr$ in state $q_\ttr$.)
We say that $S$ is {\em reception-error free} whenever, for any $s\in \RS(S)$, $s$ is not an unspecified reception configuration.
\item
\label{def:progress-i}
$S$ satisfies the {\em progress property} if for all $s= (\vec{q},\vec{w}) \in \RS(S)$, either there exists $s'$ such that $s\lts{} s'$
or $(\forall \ttp\in\roles. ~ q_\ttp \text{ is final})$. 
\item
\label{def:lock-freedom}
$s$ is a $\ttp$-{\em lock configuration} of $S$ if $\ttp\in\roles$,  $q_{\ttp}$ is a receiving state and \\
\centerline{$ \text{ $\ttp$ does not appear as subject in any label of any transition sequence from $s$}
$}
i.e. $\ttp$ remains stuck in all possible transition sequences from $s$.
We say that $S$ is {\em lock-free} whenever, for each $\ttp\in\roles$ and each $s\in \RS(S)$, $s$ is not a $\ttp$-lock configuration.
%\item
%%{\em \cite{TY15}}
%$S$ is {\em safe} if,  for each $s\in \RS(S)$, $s$
%is neither a deadlock, nor an orphan-message, nor an unspecified reception
%configuration.

\end{enumerate}
\end{definition}

Note that progress property (\ref{def:progress-i}) implies deadlock-freeness. 
Moreover, an unspecified reception configuration is trivially a $\ttp$-lock for some 
$\ttp$. This immediately implies that lock-freeness implies
reception-error-freeness.
It is also straightforward to check that lock-freeness does imply  both  deadlock-freeness
 and progress. 
The other properties are mutually independent. 
%\brc I believe that lock-freeness also implies progress.
%Then we have that lock-freeness implies everything but orphan-message freeness
%and the converse does also not hold. Also we know from our previous paper that
%all other properties are mutually independent which then should also hold here.
%But before we must say that  lock-freeness also implies progress if you agree.
%\erc

The above definitions of communication properties (\ref{def:safeness-i})--(\ref{def:progress-i}) are the same as the properties considered in~\cite{DY12},
though the above formulation of progress is slightly simpler but equivalent to the one in~\cite{DY12}.
The notions of orphan message and unspecified reception are also the same as in~\cite{TY15}.
The same notions of deadlock and unspecified reception are given in~\cite{CF05} and inspired by~\cite{BZ83}. The deadlock notions in~\cite{BZ83} and~\cite{TY15} coincide with~\cite{CF05} and~\cite{DY12} if the local CFSMs have no final states. Otherwise deadlock in~\cite{TY15} is weaker than deadlock above.
A still weaker notion of deadlock configuration, and hence a stronger notion of deadlock-freeness, has been suggested in~\cite{TG18}. 
This deadlock notion has been formally related to the above 
communication properties in~\cite{BdLH19}.

%\brc
%A further comment: If we can save enough space, I would so much
%prefer to move the definition of projection and~\cref{lem:nohatrestrict}
%to the main part of the paper as a hint for the most important proposition in our proofs.
%\erc

%To distinguish it from the notion above, we call it \emph{strong deadlock-freeness}, as done in \cite{BdLH19}.

%The above definitions of communication properties are the same as the properties considered in~\cite{DY12}.
%Our formulation of progress is slightly simpler but equivalent to the definition in~\cite{DY12}.
%The same notions of deadlock and unspecified reception are also given in~\cite{CF05} and inspired by~\cite{BZ83}. The deadlock notions in~\cite{BZ83} and in~\cite{TY15} coincide with~\cite{CF05} and~\cite{DY12} if the local CFSMs have no final states. Otherwise deadlock in  
%A weaker notion of deadlock configuration, and hence a stronger notion of deadlock-freeness, has been suggested in~\cite{TG18}. To distinguish it from the notion above, we call it \emph{strong deadlock-freeness}.

%!TEX root = Main-asynchCFSM-multicomp.tex
\section{PaI Multicomposition of Communicating Systems}
\label{sec:opensys}

As described in~\cref{sec:pai-multicomp}, the PaI approach to multicomposition of systems 
consists in replacing, in each to-be-composed system, one participant 
identified as an interface by a  forwarder (that we dub ``gateway'').
Any participant in a system,  say $\hh$, can be considered as an interface.
This means that we can look  at the CFSM $\hh$  
as an
abstract description of what the system expects   
from a number of ``outer'' systems (the environment) through their respective interfaces.
Hence, any message received by $\hh$ from another participant $\ttp$ of the system (to which $\hh$ belongs)
is  interpreted as a message to be forwarded to some other interface $\hh'$ among the available ones. Conversely, any message sent from $\hh$ to another participant $\ttp$
of the system (to which $\hh$ belongs)
is  interpreted as a message to be received from some other interface $\hh'$ and to be forwarded to $\ttp$.

 In order to clarify the notions introduced in this section, % PaI multicomposition for communicating systems, 
 we present below an example from \cite{BDGY23}, ``implemented'' here  in the CFSM formalism.

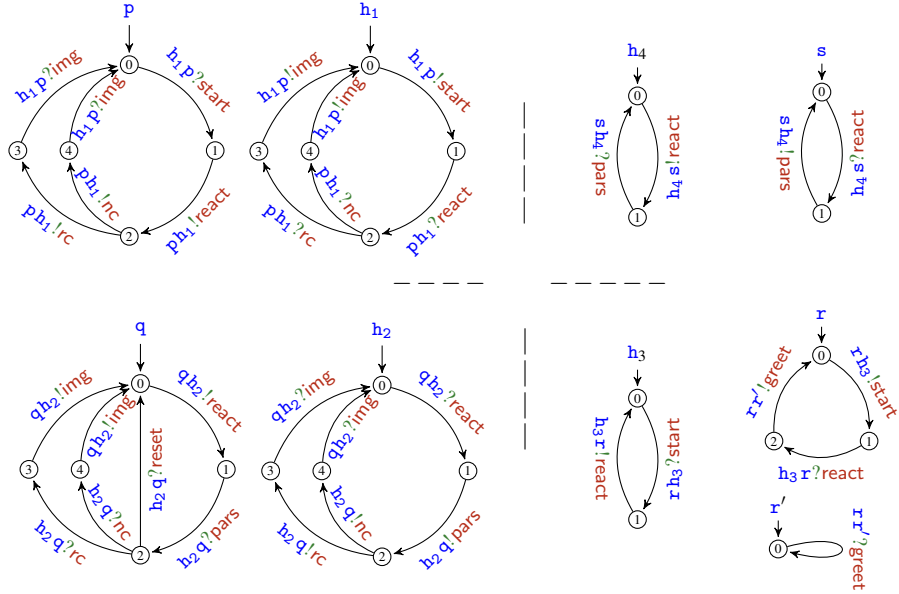
\begin{figure}[t] 
    \centering{\small
   % \vspace{-2mm}
    $
    \begin{array}{@{\hspace{-10mm}}c@{\hspace{-10mm}}}
    \begin{array}{c}
  \begin{tikzpicture}[mycfsm]
		  \node[state, initial, initial where = above, initial text={$\ptp[\ttp]$}] (0) {$0$};
		  \node[state] (1) [below right of=0]   {$1$};
		  \node[state] (2) [below left of=1]   {$2$};
		  \node[state] (3) [above left of=2,xshift=-4mm]   {$3$};
		  \node[state] (4) [above left of=2,xshift=4mm]   {$4$};
		  \path
		  (0) edge [bend left] node[above] {$\ain[h_1][p][][start]$} (1)
		  (1) edge [bend left]  node[below] {$\aout[p][h_1][][react]$} (2)
		  (2) edge [bend left]  node[below] {$\aout[p][h_1][][rc]$} (3)
		  (3) edge[bend left] node[above] {$\ain[h_1][p][][img]$} (0)
		  (2) edge [bend left]  node[above] {$\aout[p][h_1][][nc]$} (4)
		  (4) edge [bend left]  node[below] {$\ain[h_{1}][p][][img]$} (0)
		  ;
		\end{tikzpicture}
		\begin{tikzpicture}[mycfsm]
		  \node[state, initial, initial where = above, initial text={$\ptp[\HH_1]$}] (0) {$0$};
		  \node[state] (1) [below right of=0]   {$1$};
		  \node[state] (2) [below left of=1]   {$2$};
		  \node[state] (3) [above left of=2,xshift=-4mm]   {$3$};
		  \node[state] (4) [above left of=2,xshift=4mm]   {$4$};
		  \path
		  (0) edge [bend left] node[above] {$\aout[h_1][p][][start]$} (1)
		  (1) edge [bend left]  node[below] {$\ain[p][h_1][][react]$} (2)
		  (2) edge [bend left]  node[below] {$\ain[p][h_1][][rc]$} (3)
		  (3) edge[bend left] node[above] {$\aout[h_1][p][][img]$} (0)
		  (2) edge [bend left]  node[above] {$\ain[p][h_1][][nc]$} (4)
		  (4) edge [bend left]  node[below] {$\aout[h_1][p][][img]$} (0)
		  ;
		\end{tikzpicture}
 \end{array}
 \hspace{1mm}
\begin{array}{c}
 \\
 \\
| \\
| \\
|\\
|\\
\end{array}
 \hspace{4mm}
 \begin{array}{c}
\begin{tikzpicture}[mycfsm]
      %\tikzstyle{every edge}=[carrow]
      % 
      \node[state] (zero) {$0$};
      \node[state] (one) [below of=zero]   {$1$};
      \node[draw=none,fill=none] (start) [above  = 0.3cm  of zero]{$\HH_4$};
      \path
      (start) edge node {} (zero) 
      (zero) edge[bend left] node[below] {$\aout[h_4][s][][react]$} (one)
      (one) edge[bend left] node[below] {$\ain[s][h_4][][pars]$} (zero)
      ;
  \end{tikzpicture}
  \qquad\quad
      \begin{tikzpicture}[mycfsm]
      %\tikzstyle{every state}=[cnode]
      %\tikzstyle{every edge}=[carrow]
      % 
      \node[state] (zero) {$0$};
      \node[state] (one) [below of=zero]   {$1$};
            \node[draw=none,fill=none] (start) [above  = 0.3cm  of zero]{$\tts$};
      \path
      (start) edge node {} (zero) 
      (zero) edge[bend left]  node[below] {$\ain[h_4][s][][react]$} (one)
      (one) edge[bend left] node[below] {$\aout[s][h_4][][pars]$} (zero)

      ;
  \end{tikzpicture}
 \end{array}
 \\[12mm]
\hspace{24mm}- - - -    \hspace{8mm}- - - - -  \\[-8mm]
\begin{array}{@{\hspace{10mm}}c}
\\[4mm]
		\begin{tikzpicture}[mycfsm]
		  \node[state, initial, initial where = above, initial text={$\ptp[\ttq]$}] (0) {$0$};
		  \node[state] (1) [below right of=0]   {$1$};
		  \node[state] (2) [below left of=1]   {$2$};
		  \node[state] (3) [above left of=2,xshift=-4mm]   {$3$};
		  \node[state] (4) [above left of=2,xshift=4mm]   {$4$};
		  \path
		  (0) edge [bend left] node[above] {$\aout[q][h_2][][react]$} (1)
		  (1) edge [bend left]  node[below] {$\ain[h_2][q][][pars]$} (2)
		  (2) edge [bend left]  node[below] {$\ain[h_2][q][][rc]$} (3)
		  (3) edge[bend left] node[above] {$\aout[q][h_2][][img]$} (0)
		  (2) edge [bend left]  node[above] {$\ain[h_2][q][][nc]$} (4)
		  (4) edge [bend left]  node[below] {$\aout[q][h_2][][img]$} (0)
		  (2) edge  node[below] {$\ain[h_2][q][][reset]$} (0)
		  ;
		\end{tikzpicture}
		\begin{tikzpicture}[mycfsm]
		  \node[state, initial, initial where = above, initial text={$\ptp[\HH_2]$}] (0) {$0$};
		  \node[state] (1) [below right of=0]   {$1$};
		  \node[state] (2) [below left of=1]   {$2$};
		  \node[state] (3) [above left of=2,xshift=-4mm]   {$3$};
		  \node[state] (4) [above left of=2,xshift=4mm]   {$4$};
		  \path
		  (0) edge [bend left] node[above] {$\ain[q][h_2][][react]$} (1)
		  (1) edge [bend left]  node[below] {$\aout[h_2][q][][pars]$} (2)
		  (2) edge [bend left]  node[below] {$\aout[h_2][q][][rc]$} (3)
		  (3) edge[bend left] node[above] {$\ain[q][h_2][][img]$} (0)
		  (2) edge [bend left]  node[above] {$\aout[h_2][q][][nc]$} (4)
		  (4) edge [bend left]  node[below] {$\ain[q][h_2][][img]$} (0)
		  ;
		\end{tikzpicture}
 \end{array}
 \hspace{-0.5mm}
\begin{array}{c}
 \\[-12mm]
| \\
| \\
| \\
|
\end{array}
 \hspace{4mm}
 \begin{array}{c}
 \\[-2mm]
   \begin{tikzpicture}[mycfsm]
      %\tikzstyle{every state}=[cnode]
      %\tikzstyle{every edge}=[carrow]
      % 
      \node[state] (zero) {$0$};
      \node[state] (one) [below of=zero]   {$1$};
      \node[draw=none,fill=none] (start) [above  = 0.3cm  of zero]{$\HH_3$};
      \path
      (start) edge node {} (zero) 
      (zero) edge[bend left]  node[below] {$\ain[r][h_3][][start]$} (one)
      (one) edge[bend left] node[below] {$\aout[h_3][r][][react]$} (zero)
      ;
  \end{tikzpicture}
  \quad
  \begin{array}{c}
  \\[2mm]
      \begin{tikzpicture}[mycfsm]
      %\tikzstyle{every edge}=[carrow]
      % 
      \node[state] (zero) {$0$};
      \node[state] (one) [below right of=zero, xshift=-6mm]   {$1$};
      \node[state] (two) [below left  of=zero, xshift=6mm]   {$2$};
      \node[draw=none,fill=none] (start) [above  = 0.3cm  of zero]{$\ttr$};
      \path
      (start) edge node {} (zero) 
      (zero) edge[bend left] node[above] {$\aout[r][h_3][][start]$} (one)
      (one) edge[bend left] node[below] {$\ain[h_3][r][][react]$} (two)
      (two) edge[bend left] node[above] {$\aout[r][r'][][greet]$} (zero)
      ;
  \end{tikzpicture}
   \\
      \begin{tikzpicture}[mycfsm]
      %\tikzstyle{every state}=[cnode]
      %\tikzstyle{every edge}=[carrow]
      % 
      \node[state] (zero) {$0$};
      %\node[state] (one) [below of=zero]   {$1$};
      \node[draw=none,fill=none] (start) [above  = 0.3cm  of zero]{$\ttr'$};
      \path
      (start) edge node {} (zero) 
      (zero) edge[loop right,looseness=40]  node[above] {$\ain[r][\ttr'][][greet]$} (zero)
      %(zero) edge[bend right] node[below] {$\aout[h_4][r][][react]$} (one)
      ;
  \end{tikzpicture}
  \end{array}
 \end{array}
 \\[-2mm]
 \end{array}
 $
 }
 \caption{ The four communicating systems formalising the systems of Example~\ref{ex:simplewe} }
\label{eq:JK}
 \end{figure}

\begin{example}[Working example] \label{ex:simplewe}
{\em 
Let us consider the following four  systems\footnote{ For the sake of simplicity,  the example considers only systems with two  or three  participants. 
 Our definitions and results are of course 
% but all our development is 
independent of the number of participants in the single systems. }: 

\begin{description}
\item 
{\em System-1} with participants $\HH_1$ and $\ttp$.\\
Participant $\HH_1$ controls the entrance of customers in a mall (via some sensor).
As soon as a customer enters,
$\HH_1$ sends a message $\msg[start]$ to the participant $\ttp$ which controls a display for
advertisements. On receiving the start message, $\ttp$ displays a general advertising image.  
Participant 
$\ttp$ does also control a sensor detecting emotional reactions as well  as  a card reader distinguishing regular from new customers. Such information, through the messages $\msg[react]$, $\msg[rc]$ and $\msg[nc]$ is sent  to $\HH_1$. Using that information $\HH_1$ sends to $\ttp$
a customised image, depending on the kind of the customer, through message $\msg[img]$.
\item 
{\em System-2} with participants $\HH_2$ and $\ttq$.\\
 Participant $\HH_2$ controls an image display. Images are provided by participant $\ttq$
 according to some parameters sent by % with sender 
 $\HH_2$ itself and depending on  the reaction  acquired by a sensor
 driven by $\ttq$. Images are chosen also in terms of 
 the kind of customers, on the basis of their cards. Participant $\ttq$ is able to receive a 
 $\msg[reset]$ message too, even if $\HH_2$ cannot ever send it.
 \item 
{\em System-3} with participants $\HH_3$, $\ttr$ and $\ttr'$.\\
Participant $\ttr$ controls a sensor detecting the entrance of people from a door.
Once someone enters, a message  $\msg[start]$ is sent by $\ttr$ to participant $\HH_3$
which turns on a light.
The reaction of who enters, detected by a sensor driven by $\HH_3$,  is sent back to
$\ttr $ which, according to the reaction, communicates to $\ttr'$ the
greeting to be  broadcasted from the loudspeakers.
 \item 
{\em System-4}  with participants $\HH_4$ and $\tts$.\\
 Some sensors driven by Participant $\HH_4$ acquire the first reactions of people getting into
a hall adorned by several Christmas  lights. Such reactions, sent to participant $\tts$ through a message $\msg[react]$, enable $\tts$ to send to $\HH_4$ a
set of parameters $(\msg[pars])$ allowing the latter to adjust the lights of the hall.
\end{description}
The behaviours of the participants of the above systems
-- assuming an asynchronous model of communication -- can be formalised as CFSMs. So the systems above can  be formalised as the following communicating systems % CSs
\\
\centerline{$
S_1=(M_{\ttx})_{\ttx\in\Set{\HH_1,\ttp}} \quad S_2=(M_{\ttx})_{\ttx\in\Set{\HH_2,\ttq}}\quad S_3=(M_{\ttx})_{\ttx\in\Set{\HH_3,\ttr,\ttr'}} \quad S_4=(M_{\ttx})_{\ttx\in\Set{\HH_4,\tts}}
$}
as described, anticlockwise, in Figure \ref{eq:JK}.
}
 \finex
\end{example}

 {\bf Notation:} We use the following notation to denote the above set of communicating systems:  \\ $\Set{S_i}_{i\in\Set{1,2,3,4}}$ where $S_i=(M_{\ttx})_{\ttx\in\roles_i}$ with
$\roles_1=\Set{\HH_1,\ttp}$, $\roles_2=\Set{\HH_2,\ttq}$, $\roles_3=\Set{\HH_3,\ttr,\ttr'}$
and $\roles_4=\Set{\HH_4,\tts}$.\\

The composition of a set of systems relies on a selection of participants,
one for each system,  considered as interfaces.

 \begin{definition}[Interfaces]\label{def:interfaces}
 Let $\Set{S_i}_{i\in I}$ be a set of communicating systems such that, for each $i\in I$, 
 $S_i=(M_{\ttx})_{\ttx\in\roles_i}$,  where the $\roles_i$'s are pairwise disjoint. 
 A set of participants $H = \Set{\hh_i}_{i\in I}\subseteq \bigcup_{i\in I}\roles_i$
 is a {\em set of interfaces} for $\Set{S_i}_{i\in I}$ whenever,
for each $i\in  I$, $\hh_i \in \roles_i$.
 An interface $\hh_i$ has \emph{no mixed states} if the CFSM $M_{h_i}$ in $S_i$
has no mixed states. 
 \end{definition}

 \begin{example}
 {\em  We choose $\Set{\hh_i}_{i\in \Set{1,2,3,4}}$ as set of interfaces for the communicating systems of Figure~\ref{eq:JK}.\finex} 
 \end{example}

\smallskip
 We introduce now the notion of {\em connection model}\footnote{Such a notion was informally introduced in \cite{BDGY23}
in the setting of MultiParty Session Types.}, formalising what we have informally called ``choice of partners'' in~\cref{sec:pai-multicomp}.
%A connection model
A connection model is intended to  specify  the structural
(architectural)  aspects of possible ``reasonable'' connections  between interfaces of systems. 
% By imposing gateways to comply with a given connection model we rule out blatantly unreasonable compositions.
%\brc  >>> OK. DONE
%I think the last sentence needs to be a bit more elaborated.
%Therefore I would replace the last sentence by the following.
%\erc
Connection models should be provided before systems are composed since they help the
system designer to avoid blatantly unreasonable compositions.
Formally, a connection model is a set of {\em connections}, where a connection is a triple 
$(\hh, \msg[a],\hh')$ in which $\hh$ and $\hh'$ are, respectively,
interfaces of two systems, say $S$ and $S'$, and $\msg[a]$ is an input  message 
for $\hh$ and an output message for $\hh'$. 
Being $\msg[a]$ an input for $\hh$, this participant is supposed to receive $\msg[a]$ from the ``inside'' of $S$, i.e.\ from another participant of $S$.
As previously mentioned, PaI multicomposition relies on the idea
that $\msg[a]$ can be forwarded to the interface of some other system.
The connection $(\hh, \msg[a],\hh')$ hence specifies that  
$\hh'$ is one of the  possible interfaces $\msg[a]$ can be forwarded to.
This is sound since $\msg[a]$ is an output of $\hh'$, i.e. it is sent by $\hh'$
to some participant of $S'$.
%If such a connector $(\hh, \msg[a],\hh')$ is specified in a connection model,
%this expresses that the system designer
%may indeed realise such a communication when systems are composed.
The actual composition will then rely on gateways (forwarders) which comply with 
% the connections forming 
the connection model taken into account.

 \begin{definition}[Connection model]\label{def:cm}
 Let $\Set{S_i}_{i\in I}$ be a set of communicating systems  and let $H$ be a set of interfaces for it.
 
 \begin{enumerate}[i)]
% \item
% A {\em connection for} $H$ is a triple $(\hh, \msg[a],W)$ where
% $\hh\in H$ and $W\subseteq H$ and such that  
%$\hh\not\in O$,
%$\msg[a]\in\inn{M_{\hh}}$ and, for each $\ptp[w]\in W$, $\msg[a]\in\outt{M_{\hh}}$.
% $W \neq \emptyset$;
 \item
 A {\em  connection model for} $H$ is a ternary relation\  \
 $\cm\subseteq H\times\mathbb{A}_\mathfrak{U}\times H$\ \
 such that,   for each $\hh\in H$  and $\msg[a] \in \mathbb{A}_\mathfrak{U}$, 
 \begin{itemize}
 \item
 $\msg[a]\in\inn{\hh}$ implies $\exists\ \hh'\in H$ s.t. 
 $\msg[a]\in\outt{\hh'}$ and $(\hh, \msg[a],\hh')\in \cm$
 \item
 $\msg[a]\in\outt{\hh}$ implies $\exists\ \hh'\in H$ s.t.
 $\msg[a]\in\inn{\hh'}$ and
 $(\hh', \msg[a],\hh)\in \cm$
 \end{itemize}
 where  $\hh\neq\hh'$. \\
% such that $(\hh, \msg[a],\hh')\in \cm$ implies $\msg[a]\in \inn{M_{\hh}}\cap\outt{M_{\hh'}}$
% and $\hh\neq\hh'$.\\
 Elements of $\cm$ are called {\em connections}. In particular,
 $(\hh, \msg[a],\hh')\in \cm$ is called {\em connection for $\msg[a]$ (from  $\hh$ to $\hh'$)}.
 We also define $\Msgs{\cm}=\Set{\msg[a] \mid (\_, \msg[a],\_)\in \cm}$
  and assume that any message $\msg[a] \in \Msgs{\cm}$
 occurs in one of the interfaces in $H$ either as an input or as an output.  
 \item
 A connection model $\cm$ for $H$ is {\em strong\/} if, 
 %A {\em strong connection model} $\cm$ for $H$ is a connection model for $H$ such that, 
 for each $\hh\in H$  and $\msg[a] \in \mathbb{A}_\mathfrak{U}$,
 \begin{itemize}
 \item
 $\msg[a]\in\inn{\hh}$ implies $\exists!\ \hh'\in H$ s.t. $(\hh, \msg[a],\hh')\in \cm$
 \item
 $\msg[a]\in\outt{\hh}$ implies $\exists!\ \hh'\in H$ s.t. $(\hh', \msg[a],\hh)\in \cm$.
 \end{itemize}
 where  $\hh\neq\hh'$ and the unique existential quantifier 
`$\exists!$' stands %, as usual, 
for ``there exists exactly one''.
%$\ \hh'\in H$ means, as usual,  that there is exactly one $\hh'\in H$ with the required property.\\
% \begin{itemize}[--]
% \item
% $\msg[a]\in\inn{M_{\hh}}$ implies $\exists\ \hh'\in H$ s.t. $(\hh, \msg[a],\hh')\in \cm$;
% \item
% $\msg[a]\in\outt{M_{\hh}}$ implies $\exists\ \hh'\in H$ s.t. $(\hh', \msg[a],\hh)\in \cm$.
% \end{itemize}
% % $\msg[a]\in\bigcup_{\hh\in H} (\inn{M_{\hh}}\cup\outt{M_{\hh}})$, there exists 
% %a connection in $\cm$ for the message $\msg[a]$.
 \end{enumerate}
 \end{definition}
 
\noindent
Connection models can be graphically represented by diagrams,
%means of drawings,
like those used in \cref{fig:twocm}. 

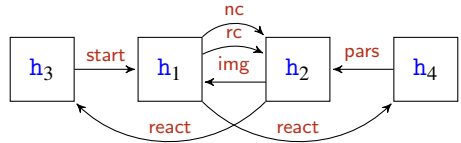
\begin{figure*}[h]\centering
 \hspace{0mm}$
\begin{tikzpicture}[mycfsm]
        \node (square-h1) at (-1,0) [draw,minimum width=1cm,minimum height=1cm] {\large $\hh_1$};
        \node (square-h2) at (1,0) [draw,minimum width=1cm,minimum height=1cm] {\large $\hh_2$};
        \node (square-h3) at (-3,0) [draw,minimum width=1cm,minimum height=1cm] {\large $\hh_3$};
       \node (square-h4) at (3,0) [draw,minimum width=1cm,minimum height=1cm] {\large $\hh_4$};
        \path   (square-h1) [bend left = 25]   edge node {$\msg[rc]$} (square-h2);
        \path (square-h4)   edge node[above]  {$\msg[pars]$} (square-h2) ;
        \path   (square-h1)[->,bend right = 45]  edge node[above] {$\msg[react]$} (square-h4)
                   (square-h2)[bend right = -45]  edge node[above] {$\msg[react]$} (square-h3)
                   (square-h3)[bend right = 0]  edge node[above] {$\msg[start]$} (square-h1)
                 ;
        \path (square-h1) [bend left = 45]  edge node {$\msg[nc]$}  (square-h2); %carrying NC from h1 to h2
        \path (0.5,-0.2)  edge node {$\msg[img]$}  (-0.5,-0.2); %carrying IMG from h1 to h2
    \end{tikzpicture}
$
   \caption{\label{fig:compsyst}A connection model for the interfaces of \cref{eq:JK}.}
\end{figure*}

\begin{example}[Some connection models]\label{ex:scm}
{\em
Let $H=\Set{\hh,\kk,\ptp[v],\ptp[w]}$ be the set of interfaces for the systems $\Set{S_i}_{i\in\Set{1,2,3,4}}$ % of Fig. \ref{fig:foursys} 
in~\cref{sec:pai-multicomp}. Fig. \ref{fig:twocm} represents the following 
connection models for $H$:
$$
\begin{array}{rcl}
\cm_{\text{A}}& = &\{ 
(\hh,\msg[a],\pw), (\pv,\msg[a],\kk),(\pw,\msg[c],\hh),(\kk,\msg[b],\pv), (\pw,\msg[b],\pv) \}\\
\cm_{\text{B}} & = &\{ 
(\hh,\msg[a],\kk), (\pv,\msg[a],\pw),(\pw,\msg[c],\hh),(\kk,\msg[b],\pv), (\pw,\msg[b],\pv)
\}
\end{array}
$$
%Of course, in general, we cannot say a connection model to be strong or not, unless we 
%either specify 
%the actual CFSMs of the  interfaces in $H$ or simply provide the information about which are the
%input and output messages of the interfaces, like graphically done in Fig. \ref{fig:foursys}.
 Obviously, both connection models %$\cm_{\text{A}}$ and $\cm_{\text{B}}$
are not strong, because of the presence of the connections $(\kk,\msg[b],\pv)$ and $(\pw,\msg[b],\pv)$. 
%\brc
%With the new def. of connection model suggested above, 
%$\cm_{\text{A}}$ and $\cm_{\text{B}}$ would not be strong since
%$(\kk,\msg[b],\pv), (\pw,\msg[b],\pv) \in \cm_{\text{A}}$ and in $\cm_{\text{B}}$. 
%\erc

 Let us now provide a connection model for the systems in Fig.\ \ref{eq:JK}
with set of interfaces $H = \Set{\hh_i}_{i\in \Set{1,2,3,4}}$. 
%For a further example of connection model, let now 
%$H = \Set{\hh_i}_{i\in \Set{1,2,3,4}}$ be the set of interfaces for the systems of Fig. \ref{eq:JK}.
First we determine $\inn{\hh_1} = \{\msg[react], \msg[nc], \msg[rc]\}$,
$\outt{\hh_1} = \{\msg[img], \msg[start]\}$,
$\inn{\hh_2} = \{\msg[react], \msg[img]\}$,\linebreak
$\outt{\hh_2} = \{\msg[nc], \msg[rc], \msg[pars]\}$,
$\inn{\hh_3} = \{\msg[start]\}$,
$\outt{\hh_3} = \{\msg[react]\}$, and
$\inn{\hh_4} = \{\msg[pars]\}$,
$\outt{\hh_4} = \{\msg[react]\}$.  \\
A connection model for $H$ is\\[1mm]
\centerline{$
\begin{array}{rcl}
\cm & = &\{ 
(\hh_1,\msg[react],\hh_4), (\hh_3,\msg[start],\hh_1),(\hh_2,\msg[img],\hh_1), 
(\hh_1,\msg[nc],\hh_2), \\
& & \hspace{2mm} (\hh_1,\msg[rc],\hh_2), (\hh_4,\msg[pars],\hh_2), 
(\hh_2,\msg[react],\hh_3)
\}
\end{array}
$}
The %graphical 
representation of $\cm$ is as in Fig. \ref{fig:compsyst}.
Obviously, this connection model is strong. 
}
\finex
 \end{example}
%\brc
%With the new def. of connection model suggested above, 
%$\cm$ would indeed be strong.
%\erc

 When we have  more than two systems to compose, the gateways are,
in general, not uniquely determined.
In order to produce  gateways out of interfaces we need to decide 
which connection model we wish to take into account and how 
the interfaces do actually interact ``complying'' with the connection model.
 Once a connection model is selected, the forwarding strategy of the gateway
is still not uniquely determined  if the connection model is not strong. 
 The reason is that in the case of at least two connectors with the same source or the same target, like $(\kk,\msg[b],\pv)$ and $(\pw,\msg[b],\pv)$ in~\cref{ex:scm}, the gateway for $\pv$
has a dynamic choice when to accept message $\msg[b]$ from $\kk$
and when from $\pw$. Therefore we need further (dynamic) information
which will be provided by {\em connection policies\/}. 
A connection policy is itself a communicating system which describes the dynamic choice of partners among the possible gateways
by respecting the constraints of (that is, complying with) the connection model.   
Technically, we first associate a set of CFSMs (the ``local connection policy set'') to each interface. 
Any element of this set specifies which communications to the ``outside'' are allowed in which state. Technically these communications are dual to the communications of its corresponding interface.

 \begin{definition}[Local Connection Policy Set]\label{def:cps}
 Let $\cm$
 be a connection model  for a set of interfaces $H$ and let $\hh\in H$
 with CFSM $M_{\HH}=(Q,q_0,\mathbb{A},\delta)$.\\
The {\em  local connection policy set} of $M_{\HH}$ w.r.t. $\cm$ 
is the set  of CFSMs  $\IS {M_\HH }{\cm}$ defined as follows:
$$ \begin{array}{ll}
\IS {M_\hh}\cm = 
\{(\dot Q,\dot{q_0},\mathbb{A},\dot\delta) \mid 
 & \dot\delta \text{  is a minimal relation s.t. $(*)$ and $(**)$}\}
  \end{array}$$
 where  $\dot Q = \Set{\dot q \mid q\in Q}$ and\\
 $(*) = 
 q \lts{\ttr\HH?\msg[a]} q'\in \delta \text{ implies } \exists \ttp\in H\setminus\Set{\hh}\, s.t.\
     \dot q \lts{\dot\HH\dot\ttp!\msg[a]} \dot{q'}\in \dot\delta \text{ and }  (\hh,\msg[a],\ttp)\in\cm$,\\
 $(**) = 
    q \lts{\HH\ttr!\msg[a]} q' \in \delta \text{ implies }  \exists \ttp\in H\setminus\Set{\hh}\, s.t.\
 \dot q \lts{\dot\ttp\dot\HH?\msg[a]} \dot{q'}\in \dot\delta \text{ and } (\ttp,\msg[a],\hh)\in\cm.$
\end{definition}

Notice that, in the above definition, each CFSM in $\IS {M_\hh}\cm$ has name $\dot \HH$.
%\brc \\To which set of participants does $\dot \HH$  belong? Probably it is a fresh name in the overall set $\roles_\mathfrak{U}$? \erc \bfc  Being a decoration of $\HH$, the name $\dot \HH$ can be definitely looked at as a fresh name. I would however refrain from discussing such an issue, in order not to 
%confuse the reader. I would leave it for a possible extended version.\\ \efc
Moreover, $\dot q$ (resp. $\dot\HH$) is to be looked at as a ``decoration'' of 
the state $q$ (resp. the name $\HH$). 
This will enable us to immediately retrieve $q$ (resp. $\HH$) out of  $\dot q$ (resp. $\dot\HH$).

\medskip
\noindent
{\bf Notation:} In the following, for the sake of readability, we shall 
write $\KK$ (resp. $\KK_i$) %as a shorthand
for $\dot\HH$ (resp. $\dot{\HH}_i$).
%i.e. $\KK$ (resp. $\KK_i$) will stand for $\dot\HH$ (resp. $\dot{\HH}_i$).\\
% refer to $\dot\HH$ (resp. $\dot{\HH}_i$)  as $\KK$ (resp. $\KK_i$).\\ 

\medskip
Local connection policy sets are finite, since they contain CFSMs % machines % processes 
which only differ in the names of participants and these names belong to a finite set.  Any element $(\dot Q,\dot{q}_0,\mathbb{A},\dot\delta)$ of $\IS {M_\hh}\cm$ 
\begin{wrapfigure}{r}{0.2\textwidth}
$
\begin{tikzpicture}[mycfsm]
		  \node[state, initial, initial where = above, initial text={$\ptp[\KK_1]$}] (0) {$\dot 0$};
		  \node[state] (1) [below right of=0]   {$\dot 1$};
		  \node[state] (2) [below left of=1]   {$\dot 2$};
		  \node[state] (3) [above left of=2,xshift=-4mm]   {$\dot 3$};
		  \node[state] (4) [above left of=2,xshift=4mm]   {$\dot 4$};
		  \path
		  (0) edge [bend left] node[above] {$\ain[k_3][k_1][][start]$} (1)
		  (1) edge [bend left]  node[below] {$\aout[k_1][k_4][][react]$} (2)
		  (2) edge [bend left]  node[below] {$\aout[k_1][k_2][][rc]$} (3)
		  (3) edge[bend left] node[above] {$\ain[k_2][k_1][][img]$} (0)
		  (2) edge [bend left]  node[above] {$\aout[k_1][k_2][][nc]$} (4)
		  (4) edge [bend left]  node[below] {$\ain[k_2][k_1][][img]$} (0)
		  ;
		\end{tikzpicture}
		\vspace{-8mm}
$
\end{wrapfigure} 
does comply with the connection model $\cm$, since it can only have transitions
$\dot q \lts{\dot\HH\ttp!\msg[a]} \dot{q'}\in \dot\delta$ with  $(\hh,\msg[a],\ttp)\in\cm$
and transitions
$\dot q \lts{\ttp\dot\HH?\msg[a]} \dot{q'}\in \dot\delta$ with $(\ttp,\msg[a],\hh)\in\cm$,
Moreover, $\IS {M_\hh}\cm$ is a singleton if the connection model $\cm$ is strong. 

\begin{example}[An element of a local connection policy set]
{\em Let $M_{\HH_1}$ be the CFSM for the participant $\HH_1$ of  \cref{ex:simplewe}
 and let $\cm$ be the  strong  connection model for $H= {\Set{\hh_i}_{i\in\Set{1,2,3,4}}}$
of Example \ref{ex:scm}. 
The CFSM on the right is the unique element of $\IS {M_{\hh_1}}\cm$.%\\[-2mm] 
\finex}
\end{example}

Given a connection model, a connection policy is obtained by choosing, for each interface, an element of its local connection policy set.
%the corresponding connection policy set .
%Of course one cannot expect an  arbitrary   
%interfacing policy to lead to a sound/reasonable composition.
%Let us consider, for example, the sessions of 
%Example~\ref{simplewe4} and an interfacing policy where we choose the following element of 
%%$\GS{\PH_1}{\set{\hh_2,\hh_3,\hh_4}}$:
%%\Cline{ \PK = 
%%\hh_2 ?\msg{start}.\,\hh_3!\msg{react}.\hh_4 !
%%\left\{ \begin{array}{l}                                                                                                                                           \msg{rc}.\,\hh_2 ?\msg{img}.\,\PK                                                                                                                               \\
%%\msg{nc}.\,\hh_2 ?\msg{img}.\,\PK
%%          \end{array}\right.}
%This would lead to a composition where the gateway we substitute for $\hh_1$
%would first expect from $\hh_2$ the message $\msg[start]$ to be forwarded to $\ttp$.
%Such a composition would immediately get stuck, since no message  $\msg[start]$
%is ever handled by $\HH_2$ and hence by the gateway we would substitute for it.
%Sound compositions will actually be the one induced by typable interfacing policies, which we dub as ``valid''.

%\begin{definition}[Connection policy]
%\label{def:connpol}
%Let $\Set{M_{\hh_i}}_{i\in I}$ be a set of CFSMs.
% A connection policy for $\Set{M_{\hh_i}}_{i\in I}$ is a CS 
% $\cp =(M_{\kk_i})_{i\in I}$ 
% such that, for each $j\in I$, $M_{\kk_j}\in \IS {M_{\hh_i}}{\Set{\kk_i}_{i\in I\setminus\Set{j}}}$.
%\end{definition}

\begin{definition}[Connection policy]
\label{def:connpol}
Let $\Set{S_i}_{i\in I}$ be a set of communicating systems such that, for each $i\in I$, 
 $S_i=(M_{\ttx})_{\ttx\in\roles_i}$, and let
$\cm$ be a connection model for a set of interfaces 
$H= \Set{\hh_i}_{i\in I}$.
 A {\em connection policy (for $H$) complying with $\cm$}  is a communicating system \
 $\cp =(M_{\kk_i})_{i\in I}$ 
 such that, for each $i\in I$, $M_{\kk_i}\in \IS {M_{\hh_i}}{\cm}$.
\end{definition}
\smallskip

Connection policies are made of local connection policies which, due to the conditions
$(*)$ and $(**)$ in \cref{def:cps}, fit to %are compliant with 
the given communication model $\cm$.
Consequently, in the above definition, the connection policy is said to be compliant with $\cm$.
If we dropped the two requirements $(*)$ and $(**)$ in \cref{def:cps} we would get non-compliant connection policies.

\smallskip
\begin{example}[A connection policy]
\label{ex:aconnpol}
{\em
The following four CFSMs constitute
a connection policy for $H=\Set{\hh_i}_{i\in I}$  complying with $\cm$, where the $M_{\hh_i}\!$'s
are as in Figure \ref{eq:JK} and $\cm$ is the  connection model  of Example \ref{ex:scm}. 
\vspace{-4mm}
{\small
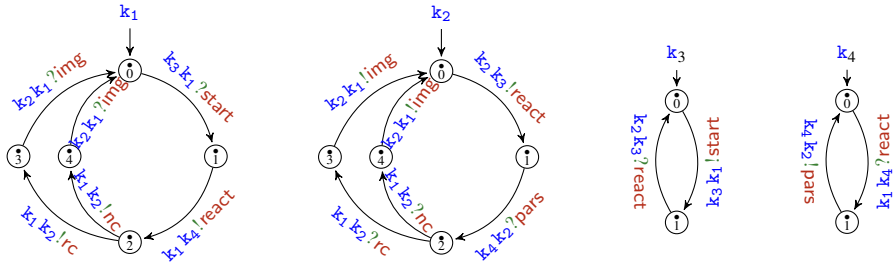
    $$
    \begin{array}{c@{\hspace{5mm}}c@{\hspace{5mm}}c@{\hspace{5mm}}c}
    \begin{array}{c}
		\begin{tikzpicture}[mycfsm]
		  \node[state, initial, initial where = above, initial text={$\ptp[\KK_1]$}] (0) {$\dot 0$};
		  \node[state] (1) [below right of=0]   {$\dot 1$};
		  \node[state] (2) [below left of=1]   {$\dot 2$};
		  \node[state] (3) [above left of=2,xshift=-4mm]   {$\dot 3$};
		  \node[state] (4) [above left of=2,xshift=4mm]   {$\dot 4$};
		  \path
		  (0) edge [bend left] node[above] {$\ain[k_3][k_1][][start]$} (1)
		  (1) edge [bend left]  node[below] {$\aout[k_1][k_4][][react]$} (2)
		  (2) edge [bend left]  node[below] {$\aout[k_1][k_2][][rc]$} (3)
		  (3) edge[bend left] node[above] {$\ain[k_2][k_1][][img]$} (0)
		  (2) edge [bend left]  node[above] {$\aout[k_1][k_2][][nc]$} (4)
		  (4) edge [bend left]  node[below] {$\ain[k_2][ k_1][][img]$} (0)
		  ;
		\end{tikzpicture}
 \end{array}
&
 \begin{array}{c}
		\begin{tikzpicture}[mycfsm]
		  \node[state, initial, initial where = above, initial text={$\ptp[\KK_2]$}] (0) {$\dot 0$};
		  \node[state] (1) [below right of=0]   {$\dot 1$};
		  \node[state] (2) [below left of=1]   {$\dot 2$};
		  \node[state] (3) [above left of=2,xshift=-4mm]   {$\dot 3$};
		  \node[state] (4) [above left of=2,xshift=4mm]   {$\dot 4$};
		  \path
		  (0) edge [bend left] node[above] {$\aout[k_2][k_3][][react]$} (1)
		  (1) edge [bend left]  node[below] {$\ain[k_4][k_2][][pars]$} (2)
		  (2) edge [bend left]  node[below] {$\ain[k_1][k_2][][rc]$} (3)
		  (3) edge[bend left] node[above] {$\aout[k_2][k_1][][img]$} (0)
		  (2) edge [bend left]  node[above] {$\ain[k_1][k_2][][nc]$} (4)
		  (4) edge [bend left]  node[below] {$\aout[k_2][k_1][][img]$} (0)
		  ;
		\end{tikzpicture}
 \end{array}
&
\begin{array}{c}
   \begin{tikzpicture}[mycfsm]
      %\tikzstyle{every state}=[cnode]
      %\tikzstyle{every edge}=[carrow]
      % 
      \node[state] (zero) {$\dot 0$};
      \node[state] (one) [below of=zero]   {$\dot 1$};
      \node[draw=none,fill=none] (start) [above  = 0.3cm  of zero]{$\KK_3$};
      \path
      (start) edge node {} (zero) 
      (zero) edge[bend left]  node[below] {$\aout[k_3][k_1][][start]$} (one)
      (one) edge[bend left] node[below] {$\ain[k_2][k_3][][react]$} (zero)
      ;
  \end{tikzpicture}
 \end{array}
 &
  \begin{array}{c}
     \begin{tikzpicture}[mycfsm]
      %\tikzstyle{every edge}=[carrow]
      % 
      \node[state] (zero) {$\dot 0$};
      \node[state] (one) [below of=zero]   {$\dot 1$};
      \node[draw=none,fill=none] (start) [above  = 0.3cm  of zero]{$\KK_4$};
      \path
      (start) edge node {} (zero) 
      (zero) edge[bend left] node[below] {$\ain[k_1][k_4][][react]$} (one)
      (one) edge[bend left] node[below] {$\aout[k_4][k_2][][pars]$} (zero)
      ;
      \end{tikzpicture}
 \end{array}
 \end{array}
 $$
% \vspace{-4mm}
 }\finex
 }
\end{example}

\begin{remark}{\em
A connection model can be looked at as a static and abstract description of connection policies. 
In particular a connection model abstracts from the order of  exchanged messages. % and from branching.
 As already pointed out above
there may be several connection policies complying with a given connection model $\cm$ if $\cm$ is not strong.
As an example assume given
three systems with the following interfaces:\\[1mm]
\centerline{$
    \centering{\small
   % \vspace{-2mm}
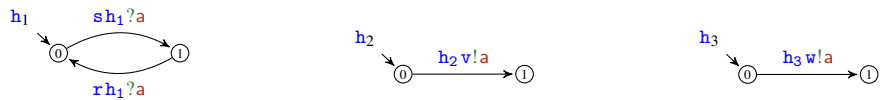
\begin{tikzpicture}[mycfsm]
      %\tikzstyle{every edge}=[carrow]
      % 
      \node[state] (zero) {$0$};
      \node[state] (one) [right of=zero]   {$1$};
      \node[draw=none,fill=none] (start) [ above left = 0.3cm  of zero]{$\HH_1$};
      \path
      (start) edge node {} (zero) 
      (zero) edge[bend left] node[above] {$\ain[s][h_1][][a]$} (one)
      (one) edge[bend left] node[below] {$\ain[r][h_1][][a]$} (zero)
      ;
  \end{tikzpicture}
 \qquad\qquad\qquad
   \begin{tikzpicture}[mycfsm]
      %\tikzstyle{every state}=[cnode]
      %\tikzstyle{every edge}=[carrow]
      % 
      \node[state] (zero) {$0$};
      \node[state] (one) [right of=zero]   {$1$};
      \node[draw=none,fill=none] (start) [ above  left = 0.3cm  of zero]{$\HH_2$};
      \path
      (start) edge node {} (zero) 
      (zero) edge  node[above] {$\aout[h_2][v][][a]$} (one)
      ;
  \end{tikzpicture}
  \qquad\qquad\qquad
   \begin{tikzpicture}[mycfsm]
      %\tikzstyle{every state}=[cnode]
      %\tikzstyle{every edge}=[carrow]
      % 
      \node[state] (zero) {$0$};
      \node[state] (one) [right of=zero]   {$1$};
      \node[draw=none,fill=none] (start) [ above left = 0.3cm  of zero]{$\HH_3$};
      \path
      (start) edge node {} (zero) 
      (zero) edge  node[above] {$\aout[h_3][w][][a]$} (one)
      ;
  \end{tikzpicture}
 }
 $}
 \noindent
 We can now consider the following (non-strong) connection model:\
 $
 \cm = \Set{(\hh_1,\msg[a],\hh_2), (\hh_1,\msg[a],\hh_3)}.
 $\\
 It is easy to check that the connection policies $\cp_1$ and $\cp_2$ below
do both comply with $\cm$.\\
\centerline{
 $
    \centering{\small
   % \vspace{-2mm}
   \text{\small $\cp_1$ = \quad}
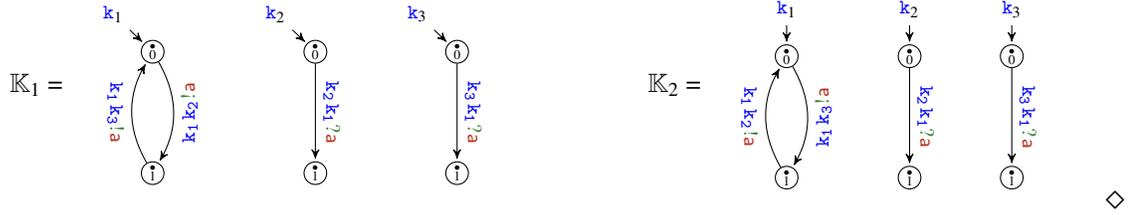
\begin{tikzpicture}[mycfsm]
      %\tikzstyle{every edge}=[carrow]
      % 
      \node[state] (zero) {$\dot 0$};
      \node[state] (one) [below of=zero]   {$\dot 1$};
      \node[draw=none,fill=none] (start) [above left = 0.3cm  of zero]{$\kk_1$};
      \path
      (start) edge node {} (zero) 
      (zero) edge[bend left] node[below] {$\aout[k_1][\kk_2][][a]$} (one)
      (one) edge[bend left] node[below] {$\aout[k_1][\kk_3][][a]$} (zero)
      ;
  \end{tikzpicture}
 \qquad
   \begin{tikzpicture}[mycfsm]
      %\tikzstyle{every state}=[cnode]
      %\tikzstyle{every edge}=[carrow]
      % 
      \node[state] (zero) {$\dot 0$};
      \node[state] (one) [below of=zero]   {$\dot 1$};
      \node[draw=none,fill=none] (start) [above left = 0.3cm  of zero]{$\kk_2$};
      \path
      (start) edge node {} (zero) 
      (zero) edge  node[above] {$\ain[k_2][k_1][][a]$} (one)
      ;
  \end{tikzpicture}
  \qquad
   \begin{tikzpicture}[mycfsm]
      %\tikzstyle{every state}=[cnode]
      %\tikzstyle{every edge}=[carrow]
      % 
      \node[state] (zero) {$\dot 0$};
      \node[state] (one) [below of=zero]   {$\dot 1$};
      \node[draw=none,fill=none] (start) [above left = 0.3cm  of zero]{$\kk_3$};
      \path
      (start) edge node {} (zero) 
      (zero) edge  node[above] {$\ain[k_3][k_1][][a]$} (one)
      ;
  \end{tikzpicture}
  \qquad\qquad\qquad
 \text{\small $\cp_2$ = \quad}
\begin{tikzpicture}[mycfsm]
      %\tikzstyle{every edge}=[carrow]
      % 
      \node[state] (zero) {$\dot 0$};
      \node[state] (one) [below of=zero]   {$\dot 1$};
      \node[draw=none,fill=none] (start) [above  = 0.3cm  of zero]{$\kk_1$};
      \path
      (start) edge node {} (zero) 
      (zero) edge[bend left] node[below] {$\aout[k_1][\kk_3][][a]$} (one)
      (one) edge[bend left] node[below] {$\aout[k_1][\kk_2][][a]$} (zero)
      ;
  \end{tikzpicture}
 \qquad
   \begin{tikzpicture}[mycfsm]
      %\tikzstyle{every state}=[cnode]
      %\tikzstyle{every edge}=[carrow]
      % 
      \node[state] (zero) {$\dot 0$};
      \node[state] (one) [below of=zero]   {$\dot 1$};
      \node[draw=none,fill=none] (start) [above  = 0.3cm  of zero]{$\kk_2$};
      \path
      (start) edge node {} (zero) 
      (zero) edge  node[above] {$\ain[k_2][k_1][][a]$} (one)
      ;
  \end{tikzpicture}
  \qquad
   \begin{tikzpicture}[mycfsm]
      %\tikzstyle{every state}=[cnode]
      %\tikzstyle{every edge}=[carrow]
      % 
      \node[state] (zero) {$\dot 0$};
      \node[state] (one) [below of=zero]   {$\dot 1$};
      \node[draw=none,fill=none] (start) [above  = 0.3cm  of zero]{$\kk_3$};
      \path
      (start) edge node {} (zero) 
      (zero) edge  node[above] {$\ain[k_3][k_1][][a]$} (one)
      ;
  \end{tikzpicture}
 }
 $}
 \finex
 }
\end{remark}

%\brc
%The next fact would fit directly after the definition of connection policy.
%\erc
%
%\bfr
%It is immediate to check the following fact.
%\begin{fact}
%For any $\hh\in H$ and any two connection models $\cm_1$ and $\cm_2$ for $H$,
%$$\cm_1\subseteq\cm_2 \implies \IS {M_{\hh}}{\cm_1} \subseteq \IS {M_{\hh}}{\cm_2}$$
%\end{fact}
%\efr 

By now we have almost all the necessary notions to formally define the PaI multicomposition of systems of communicating systems. The only missing piece is that of
building the gateways using a connection policy.

We get a gateway essentially by transforming an interface $M_\HH$ by inserting  a fresh state in between any transition. 
Any input  transition  $q\lts{\tts\hh?\msg[a]}q'$ (resp. output transition $q\lts{\hh\tts!\msg[a]}q'$) of $M_\HH$ is then transformed into two consecutive transitions \\
\centerline{
$q\lts{\tts\hh?\msg[a]}\hat{q} \lts{\hh\hh'!\msg[a]}q'\qquad 
\text{(resp. $q\lts{\hh'\hh?\msg[a]}\hat{q} \lts{\hh\tts!\msg[a]}q'$)}$}
where $\hat{q}$ is a fresh state and 
$\dot{q}\lts{\kk\kk'!\msg[a]}\dot{q'}\,\,\
\text{(resp. $q\lts{\kk'\kk?\msg[a]}\dot{q'}$)}$
belonging to 
% $(\hh,\msg[a],\hh')$ is a connection 
%one of the other interfaces, according 
the connection policy taken into account.
In the formal definition below we distinguish the fresh  states by superscripting them
by the transition they are ``inserted in between''.

%In such a way a transition from $q$ to $q'$ receiving a message $\msg[a]$ from a role $\tts$ is transformed
%into two transitions: one from $q$ to the new state $\widehat{q}$ receiving $\msg[a]$ from $\tts$, and one from
%$\widehat{q}$ to $q'$ sending $\msg[a]$ to one of the other interfaces, according to the connection policy used. 
%Conversely, a transition from $q$ to $q'$ sending a message $\msg[a]$ to a role $\tts$ is transformed
%into two transitions: one from $q$ to the new state $\widehat{q}$ receiving $\msg[a]$ from an interface (according to the connection policy), and one from
%$\widehat{q}$ to $q'$ sending $\msg[a]$ to $\tts$. We distinguish the new ``inserted'' states by superscripting them
%by the transition where they are ``inserted in between''.

\begin{definition}[Gateway]\hfill\\
\label{def:gatewaymc} Assume given a connection model $\cm$ 
and two %no-mixed-state\footnote{This condition is necessary in order to get a sound definition of initial state for the gateway, and it is however among those needed to get a safe composition (see  \cref{def:composability} and subsequent observation).} 
CFSMs $M_{\hh}$ and $M_{\kk}$ such that 
 $M_{\hh}= (Q, q_0,\mathbb{A},\delta)$ and 
$M_{\kk} = (\dot Q,\dot {q_0},\mathbb{A},\dot\delta)\in \IS {M_{\hh}}{\cm}$. 
The {\em gateway} $M_{\hh}{\gts} M_{\kk}$ obtained out of  $M_\HH$ and $M_\KK$  is defined by  
$$
M_{\hh}{\gts} M_{\kk} = (Q\cup\widehat Q, q_0, \mathbb{A},\hat{\delta})
$$
where\\
$-$ $\widehat{Q} =\bigcup_{q\in Q}\Set{q^{(q, l,q')} \mid (q, l,q')\in\delta},$ %\bigcup_{q\in Q}\Set{q^{(q,\msg[a],q')} \mid (q,\msg[a],q')\in\delta}
\\
$-$ $\widehat\delta = \Set{(q,\ttr\HH?\msg[a],\widehat q), (\widehat q,\HH\tts!\msg[a],q') \mid  (q,\HH\tts!\msg[a],q')\in\delta, (\dot q,\dot\ttr\KK?\msg[a],\ \dot{q'})\in\dot\delta,\ \widehat q=q^{(q,\HH\tts!\msg[a],q')}}$\\
${\qquad}\cup\Set{(q,\tts\HH?\msg[a],\widehat q), (\widehat q,\HH\ttr!\msg[a],q') \mid  (q,\tts\HH?\msg[a],q')\in\delta,\ (\dot q,\KK\dot\ttr!\msg[a],\dot{q'})\in\dot\delta,\ \widehat q=q^{(q,\tts\HH?\msg[a],q')}}.$

%$$
%M_{\hh}{\gts} M_{\kk} = (Q\cup\widehat Q, \ddot{q}_0,\mathbb{A},\hat{\delta})
%$$
%where\\
%\brc
%I made a mess here and cannot undo it.
%Here was the definition of  $\ddot{q}_0$ by a case distinction:\\
%$\ddot{q}_0 = q_0 \text{ if the outgoing transitions from  $q_0$  are input transitions}$\\
%$\ddot{q}_0 = \dot{q}_0 \text{ if the outgoing transitions from  $q_0$  are output transitions}$\\
%Anyway, I cannot understand why $\ddot{q}_0 = \dot{q}_0$ in the second case. $\dot{q}_0$ is not an element of  $Q\cup\widehat Q$.\\
%I would expect that always $\ddot{q}_0 = q_0$.
%\erc\\
%%$-$ $\ddot{q}_0 = \left\{ \begin{array}{ll}
%%                                       q_0 &  \text{ if the outgoing transitions from  $q_0$  are input transitions}\\
%%                                       \dot{q}_0 &  \text{ if the outgoing transitions from  $q_0$  are output transitions}                       
%%                                     \end{array}
%%                            \right
%%$\\
%$-$ $\widehat{Q} =\bigcup_{q\in Q}\Set{q^{(q, l,q')} \mid (q, l,q')\in\delta}$ %\bigcup_{q\in Q}\Set{q^{(q,\msg[a],q')} \mid (q,\msg[a],q')\in\delta}
%\\
%$-$ $\widehat\delta = \Set{(q,\ttr\HH?\msg[a],\widehat q), (\widehat q,\HH\tts!\msg[a],q') \mid  (q,\HH\tts!\msg[a],q')\in\delta, (\dot q,\dot\ttr\KK?\msg[a],\ \dot{q}')\in\dot\delta,\ \widehat q=q^{(q,\HH\tts!\msg[a],q')}}$\\
%${\qquad}\cup\Set{(q,\tts\HH?\msg[a],\widehat q), (\widehat q,\HH\ttr!\msg[a],q') \mid  (q,\tts\HH?\msg[a],q')\in\delta,\ (\dot q,\KK\dot\ttr!\msg[a],\dot{q}')\in\dot\delta,\ \widehat q=q^{(q,\tts\HH?\msg[a],q')}}$

\smallskip
 
\noindent
We refer to $\widehat\delta$ as $\widehat\delta_{\HH}$ whenever $\HH$ is not clear from the
context; similarly for $\widehat Q$.
\end{definition}

\begin{example}[A gateway]{\em 
Let $M_{\HH_1}$ be as in Example \ref{ex:simplewe},
and let $M_{\KK_1}$ be as in %the CFSM for participant $\KK_1$ in 
the connection policy of \cref{ex:aconnpol}.
The gateway $M_{\HH_1}\gts M_{\KK_1}$ is as follows.\\
\centerline{$
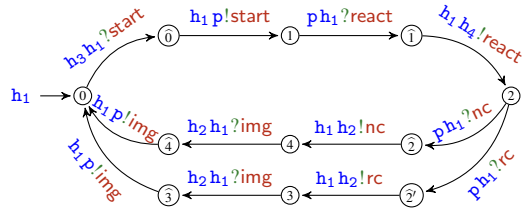
\begin{tikzpicture}[mycfsm]
		  \node[state, initial, initial where = left, initial text={$\ptp[\HH_1]$}] (0) {$0$};
		  \node[state] (hat0) [above right of=0,yshift=-4mm]   {$\widehat 0$};
		  \node[state] (1) [right of=hat0]   {$1$};
		  \node[state] (hat1) [right of=1]   {$\widehat 1$};
		  \node[state] (2) [right of=0, xshift=48mm]   {$2$};
		  \node[state] (hat2) [below of=hat1, yshift=2mm]   {$\widehat 2$};
		  \node[state] (hat2p) [below of=hat1, yshift=-6mm]   {$\widehat{2}'$};
		  \node[state] (3) [below of=1,yshift=-6mm]   {$3$};
		  \node[state] (4) [below of=1, yshift=2mm  ] {$4$};
		  \node[state] (hat3) [below of=hat0,yshift=-6mm]   {$\widehat 3$};
		  \node[state] (hat4) [below of=hat0, yshift=2mm  ] {$\widehat 4$};
		  \path
		  (0) edge [bend left] node[above] {$\ain[h_3][h_1][][start]$} (hat0)
		  (hat0) edge node[above] {$\aout[h_1][p][][start]$} (1)
		  (1) edge   node[above] {$\ain[p][h_1][][react]$} (hat1)
		  (hat1) edge  [bend left] node[above] {$\aout[h_1][h_4][][react]$} (2)
		  (2) edge [bend left]  node[below] {$\ain[p][h_1][][rc]$} (hat2p)
		  (hat2p) edge   node[above] {$\aout[h_1][h_2][][rc]$} (3)
		  (3) edge node[above] {$\ain[h_2][h_1][][img]$} (hat3)
		  (hat3) edge[bend left] node[below] {$\aout[h_1][p][][img]$} (0)
		  (2) edge [bend left]  node[above] {$\ain[p][h_1][][nc]$} (hat2)
		  (hat2) edge   node[above] {$\aout[h_1][h_2][][nc]$} (4)
		  (4) edge node[above] {$\ain[h_2][h_1][][img]$} (hat4)
		  (hat4) edge[bend left] node[above] {$\aout[h_1][p][][img]$} (0)
		  ;
		\end{tikzpicture}
\vspace{-4mm}
$}
\finex
}\end{example}

%In order to get a safe composition, some conditions are needed.

 \begin{definition}[Composability]
 \label{def:composability}
Let $\Set{S_i}_{i\in I}$ be a set of communicating systems such that, for each $i\in I$, $S_i=(M_{\ttx})_{\ttx\in\roles_i}$. 
Moreover, let $H=\Set{\hh_i}_{i\in I}$ be a set of interfaces for it.
We say that $\Set{S_i}_{i\in I}$ is {\em composable with respect to} $H$ whenever
the  sets $\roles_i$'s are  pairwise disjoint.
%\begin{enumerate}[a)]
%\item
%the  $\roles_i$'s are  pairwise disjoint;
%\item
%the $M_{\hh_i}$'s are ?!-deterministic and have no mixed states.
%\brc
%Probably we can remove no-mixed-state also and add it as an assumption
%for preservation of no unspecified reception.
%\erc
%\end{enumerate}
\end{definition} 

%\begin{remark}\em 
%One could wonder why, for the property preservation results of~\cite{BdLH19} for the binary case,
%additional conditions like $?!$-determinism (see ~\cite[Sect.2]{BdLH19}) are required for interfaces. 
%As a matter of fact, in the present paper we are assuming that communication properties do
%hold for the communication policies, whereas in~\cite{BdLH19} compatibility is assumed.
%The latter is not enough to ensure communication properties for the communication policies
%unless further conditions like $?!$-determinism are also considered. 
%\finex
%\end{remark}

%
%PaI multicomposition fails to be safe without the above conditions, even for the binary case.
%We refer to counterexamples in \cite[Sect.5]{BdLH19} for the necessity of such conditions.

Let us now describe how systems are composed on the basis of a given connection policy.
%for their interfaces.

\begin{definition}[Multicomposition of communicating systems]
\label{def:multicomposition}
%Let $\Set{S_i}_{i\in I}$ be a set of CSs such that, for each $i\in I$, $S_i=(M_\ttp)_{\ttp\in\roles_i}$.
%Let $H=\Set{\hh_i}_{i\in I}$ be a set of interfaces such that
%$\Set{S_i}_{i\in I}$ is composable with respect to $H$,
%and let $\cm$ be a connection model for 
%$H.$ Moreover, let $\cp=(M_{\kk_i})_{i\in I}$ be a connection policy for $\Set{M_{\hh_i}}_{i\in I}$
% complying with $\cm$. 
Let $\Set{S_i}_{i\in I}$ be a set of communicating systems composable with respect to $H=\Set{\hh_i}_{i\in I}$ 
and let $\cp=(M_{\kk_i})_{i\in I}$ be a connection policy
 complying with a connection model $\cm$ for $H$. 
The {\em multicomposition of $\Set{S_i}_{i\in I}$ with respect to $\cp$} 
%(dubbed $\MC(\Set{S_i}_{i\in I}, \cp)$)
 is the communicating system 
$$\MC(\Set{S_i}_{i\in I}, \cp) =  (M'_\ttp)_{\ttp\in\bigcup_{i\in I}\roles_i}$$
where\\
${\qquad\qquad}M'_\ttp = \left\{ \begin{array}{ll}
                          M_\ttp & \text{ if } \ttp\not\in\Set{\hh_i}_{i\in I}\\[2mm]
                          M_{\hh_i}{\gts\,}M_{\kk_i} & \text{ if } \ttp=\hh_i \text{ with $i\in I$}
                           \end{array}
                 \right.$
\end{definition}

\smallskip
Note that the CFSMs of a composition are CFSMs over $\roles = \bigcup_{i\in I}\roles_i$
 and $\mathbb{A} = \bigcup_{i\in I} \mathbb{A}_i$.
 Graphically, the architectural structure of a multicomposition via gateways can be shown as in~\cref{fig:multiconnection}.

%\begin{remark}
%{\em
%Our forthcoming result about safety of  multicomposition is actually independent of a  concrete connection model.
%Considering connection policies which comply with a connection model is, however,
%helpful at the design stage of the multicomposition and
%enhances the possibility of getting connection policies which 
%satisfy communication properties and hence support the preservation of communication properties of the composed systems.
%\finex
%}
%\end{remark}

\section{On the Preservation of Communication Properties}
\label{sec:preservation}

The main result of the present paper is the safety of PaI multicomposition of CFSM 
systems for all communication properties of~\cref{def:safeness} but lock-freeness.
Apart from orphan-message-freeness we need the no-mixed-state assumption for interfaces to obtain the preservation results.

\begin{theorem}[Safety of PaI multicomposition of CFSM systems]
\label{th:paisafenesse}
 Let $\Set{S_i}_{i\in I}$ be a set of communicating systems composable with respect to a set
 $H = \{\hh_i\}_{i \in I}$
 of interfaces with no mixed states (cf.~\cref{def:interfaces}) and
let $\cp$ be a connection policy for $H$. 
Let $\mathcal{P}$ be
either the property of {\em deadlock-freeness} or
{\em reception-error-freeness} %or {\em strong deadlock freeness}
or {\em progress}.
%(as defined in \cref{def:safeness}).
%\centerline{
If $\mathcal{P}$ holds for each $S_i$ with $i \in I$ %CS in $\Set{S_i}_{i\in I}$
and for $\cp$, 
then $\mathcal{P}$ holds for $S  = \MC(\Set{S_i}_{i\in I}, \cp)$.
%}
Moreover, the above holds also if the no-mixed-state condition is removed and
$\mathcal{P}$ is {\em orphan-message-freeness}.
\end{theorem}

\begin{remark}
{\em
 The above %The Our forthcoming 
result about safety of  multicomposition is actually independent of a  concrete connection model.
Considering connection policies which comply with a connection model is, however,
helpful at the design stage of the multicomposition and
enhances the possibility of getting connection policies which 
satisfy communication properties and hence support the preservation of communication properties of the composed systems.
\finex
}
\end{remark}

%This theorem 
 \cref{th:paisafenesse} can be proved for each property $\mathcal{P}$ separately by contradiction. 
 In particular by showing that $\mathcal{P}$ does not hold for $S$ implies that it does not hold either for one of the $S_i$'s or for $\cp$. 

A key notion for the proofs is that of {\em  projection\/} of a reachable configuration of the composed system to configurations of each of the single systems $S_i$ and also of the connection policy $\cp$. On this basis, the most important tool to get contradictions is the subsequent~\cref{lem:nohatrestrict-maintext} which essentially shows that projections of reachable configurations involving no intermediate gateway states are reachable configurations again. %of the single systems and of $\cp$.
The complete proofs of property preservations are provided  in~\cite{BH24full}.
%\cref{sect:safetypreservation}. 
They are independent of the communication model $\cp$ complies with. 

%We shall prove that if a projection does not contain any fresh state
%introduced by gateway construction then it is a configuration in the corresponding system.
 
\begin{definition}[Configuration projections]
\label{def:projectedconf-maintext}
 Let $S = \MC(\Set{S_i}_{i\in I}, \cp)$ be as in~\cref{th:paisafenesse}
(but without no-mixed-state assumption).
%$\cp =(M_{\kk_i})_{i\in I}$ is a connection policy for $\Set{M_{\hh_i}}_{i\in I}$.
% Moreover, 
Let $s= (\vec{q},\vec{w})\in \RS(S)$ where $\vec{q}=(q_\ttp)_{\ttp\in\roles}$
and $\vec{w} = (w_{\ttp\ttq})_{\ttp\ttq\in C_\roles}$.
 For each $i\in I$, the projection $\restrict{s}{i}$ of $s$
to $S_i$ is defined by\\
\centerline{$\restrict{s}{i}=(\restrict{\vec{q}}{i},\restrict{\vec{w}}{i})$}
where $\restrict{\vec{q}}{i} = (q_\ttp)_{\ttp\in\roles_i}$ and 
$\restrict{\vec{w}}{i} =  (w_{\ttp\ttq})_{\ttp\ttq\in C_{\roles_i}}$.

\noindent
The projection $\restrict{s}{\cp}$ of $s= (\vec{q},\vec{w})$ to $\cp$  is defined
if ${q}_{\HH_i}\not\in \widehat Q_{\HH_i}$  for each $i \in I$ and
then \\[1mm]
\centerline{$\restrict{s}{\cp}=(\restrict{\vec{q}}{\cp},\restrict{\vec{w}}{\cp})$}\\[1mm]
where $\restrict{\vec{q}}{\cp} = (p_{\kk_i})_{i\in I}$ is such that, for each $i \in I$,
 $p_{\kk_i} = \dot{q_{\HH_i}}$ (with $\dot{q_{\HH_i}}$ being the ``dotted decoration'' of the local state $q_{\HH_i}$)  
and where
$\restrict{\vec{w}}{\cp} =  (w'_{\ttp\ttq})_{\ttp,\ttq\in \Set{\KK_i}_{i\in I},\ttp\neq\ttq}$  
is such that, for each pair ${i,j\in I}$ with $i\neq j$,
$w'_{\KK_i\KK_j} = w_{\HH_i\HH_j}$.
\end{definition}

\medskip
\noindent
%Notice that $\restrict{s}{i}$ is not necessarily a configuration of $S_i$, because of the possible presence of the additional states introduced by the gateways construction.
%If reachable configurations of the composed system  $S=\MC(\Set{S_i}_{i\in I}, \cp)$
%do not involve intermediate gateway states
%%$M_\HH = \gateway{M^1_\HH, \KK}$, %$M_\KK = \gateway{M^2_\KK, \HH}$, 
%then, by projection,
%% by taking into account only the states of machines of  $S_i$  and disregarding the channels between the gateways, %(see Definition \ref{def:restrictedconf} above),
%we get reachable configurations of  $S_i$ and $\cp$.  

\begin{proposition}[On reachability of projections]
\label{lem:nohatrestrict-maintext}
Let $s= (\vec{q},\vec{w}) \in \RS(S)$. % be a reachable configuration of $S = \MC(\Set{S_i}_{i\in I}, \cp)$.
\begin{enumerate}[i)]
\item
\label{lem:nohatrestrict-a}
 For each $i\in I$, (${q}_{\HH_i}\not\in\widehat{Q_{\HH_i}} \implies 
\restrict{s}{i}\in \RS(S_i))$;
%\brc Shouldn't we use $i$ instead of $k$ as above?\erc
\item
\label{lem:nohatrestrict-b}
$({q}_{\HH_i}\not\in\widehat{Q_{\HH_i}}$ for each $i\in I)$ $\implies$
$\restrict{s}{\cp}\in \RS(\cp)$.
\end{enumerate}
\end{proposition}
\smallskip
The connection policy of \cref{ex:aconnpol} does enjoy all the properties of \cref{def:safeness}.
Moreover, the interfaces of the four systems of  \cref{ex:simplewe} are all with no mixed state.
Hence \cref{th:paisafenesse} guarantees that
any property (among those of \cref{def:safeness}, but lock-freedom) enjoyed by the systems is also enjoyed by
their PaI multicomposition.

\smallskip

Now we provide some examples for cases in which communication properties are
not preserved. First we show that all the three properties for which we have assumed
the no-mixed-state condition in~\cref{th:paisafenesse} would, in general, not be preserved by composition if the condition is dropped.
In the counterexamples, the receiving states introduced by the gateway construction cause the breaking of the property taken into account.

%\medskip
%Reception-error-freeness and deadlock-freeness are not preserved in general by composition, in case we dropped the {\em no-mixed-state} condition, as shown by the following examples.
%In both examples the receiving states introduced by the gateway construction cause the breaking of
%the property taken into account.

\begin{example}[No-mixed-state counterexample for deadlock-freeness and progress preservation]
\label{ex:lackprogdfpres}
\em
Let us consider the two following systems $S_1$ and $S_2$ with interfaces,
respectively, $\hh_1$ and $\hh_2$
containing mixed states.
$$
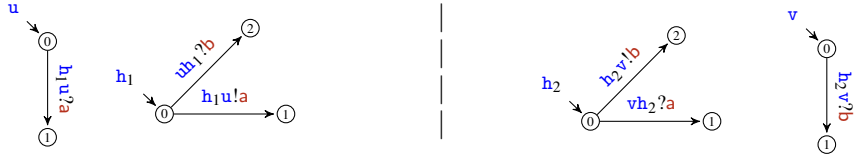
\begin{array}{c@{\qquad\qquad}c@{\hspace{1cm}}c@{\qquad}c}
    \begin{array}{cc}
      \begin{tikzpicture}[mycfsm]
  \node[state]           (0)                        {$0$};
   \node[draw=none,fill=none] (start) [above left = 0.3cm  of 0]{$\ttu$};
   \node[state]            (1) [below of=0, yshift=4mm] {$1$};

   \path  (start) edge node {} (0)
            (0)  edge    node [above] {$\hh_1\ttu?\msg[a]$} (1) ;
       \end{tikzpicture}
&
       \begin{tikzpicture}[mycfsm]
  \node[state]           (0)                        {$0$};
   \node[draw=none,fill=none] (start) [above left = 0.3cm  of 0]{$\hh_1$};
  \node[state]            (1) [right of=0] {$1$};
  \node[state]           (2) [above right of=0] {$2$};

   \path  (start) edge node {} (0) 
            (0)  edge                                   node [above] {$\hh_1\ttu!\msg[a]$} (1)
                      edge                                   node [above]  {$\ttu\hh_1?\msg[b]$} (2);
       \end{tikzpicture}
    \end{array}
       &
       \begin{array}{c}
       |\\
       |\\
       |\\
       |
       \end{array}
       &
       \begin{tikzpicture}[mycfsm]
  \node[state]           (0)                        {$0$};
   \node[draw=none,fill=none] (start) [above left = 0.3cm  of 0]{$\hh_2$};
  \node[state]            (1) [right of=0] {$1$};
  \node[state]           (2) [above right of=0] {$2$};

   \path  (start) edge node {} (0) 
            (0)  edge                                   node [above] {$\ttv\hh_2?\msg[a]$} (1)
                      edge                                   node [above]  {$\hh_2\ttv!\msg[b]$} (2);
       \end{tikzpicture}
&
      \begin{tikzpicture}[mycfsm]
  \node[state]           (0)                        {$0$};
   \node[draw=none,fill=none] (start) [above left = 0.3cm  of 0]{$\ttv$};
   \node[state]            (1) [below of=0, yshift=4mm] {$1$};

   \path  (start) edge node {} (0)
            (0)  edge    node [above] {$\hh_2\ttv?\msg[b]$} (1) ;
       \end{tikzpicture}
\end{array}
$$

$S_1$ and $S_2$ are both deadlock free and both enjoy the progress property.
There is a unique communication model for their composition:
%\centerline{
$
\cm = \Set{(\hh_2,\msg[a],\hh_1),(\hh_1,\msg[b],\hh_2)}
$
% }
The unique communication policy complying with $\cm$ is the following one.
$$
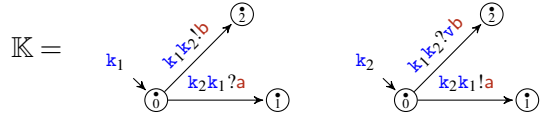
\begin{array}{c}
      \cp  = \quad
       \begin{tikzpicture}[mycfsm]
  \node[state]           (0)                        {$\dot 0$};
   \node[draw=none,fill=none] (start) [above left = 0.3cm  of 0]{$\kk_1$};
  \node[state]            (1) [right of=0] {$\dot 1$};
  \node[state]           (2) [above right of=0] {$\dot 2$};

   \path  (start) edge node {} (0) 
            (0)  edge                                   node [above] {$\kk_2\kk_1?\msg[a]$} (1)
                      edge                                   node [above]  {$\kk_1\kk_2!\msg[b]$} (2);
       \end{tikzpicture}
\qquad
       \begin{tikzpicture}[mycfsm]
  \node[state]           (0)                        {$\dot 0$};
   \node[draw=none,fill=none] (start) [above left = 0.3cm  of 0]{$\kk_2$};
  \node[state]            (1) [right of=0] {$\dot 1$};
  \node[state]           (2) [above right of=0] {$\dot 2$};

   \path  (start) edge node {} (0) 
            (0)  edge                                   node [above] {$\kk_2\kk_1!\msg[a]$} (1)
                      edge                                   node [above]  {$\kk_1\kk_2?\ttv\msg[b]$} (2);
       \end{tikzpicture}
\end{array}
$$

Also $\cp$ is deadlock free and  enjoys the progress property.  The system $\MC(\Set{S_1,S_2}, \cp)$  is the following one.
$$
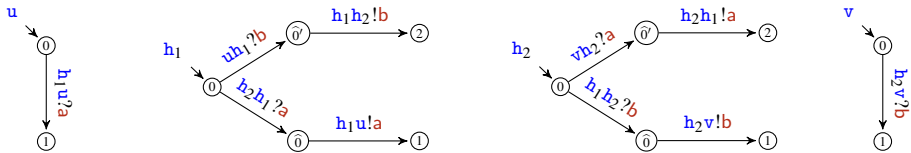
\begin{array}{c@{\hspace{1cm}}c@{\hspace{1cm}}c@{\qquad}c}
      \begin{tikzpicture}[mycfsm]
  \node[state]           (0)                        {$0$};
   \node[draw=none,fill=none] (start) [above left = 0.3cm  of 0]{$\ttu$};
   \node[state]            (1) [below of=0, yshift=4mm] {$1$};

   \path  (start) edge node {} (0)
            (0)  edge    node [above] {$\hh_1\ttu?\msg[a]$} (1) ;
       \end{tikzpicture}
&
             \begin{tikzpicture}[mycfsm]
  \node[state]           (0)                        {$0$};
  \node[state]           (hat0)          [below right of=0, yshift=5mm]              {$\widehat{0}$};
   \node[draw=none,fill=none] (start) [above left = 0.3cm  of 0]{$\HH_1$};
  \node[state]            (1) [right of=hat0] {$1$};
  \node[state]           (hat0') [above right of=0, yshift=-5mm] {$\widehat{0}'$};
  \node[state]           (2) [right of=hat0'] {$2$};

   \path  (start) edge node {} (0) 
            (0)  edge                                   node [above] {${\hh_2\hh_1}?{\msg[a]}$} (hat0)
                  edge                                   node [above]  {${\ttu\hh_1}?{\msg[b]}$} (hat0')
             (hat0)  edge                       node [above] {${\HH_1\ttu}!{\msg[a]}$} (1)
             (hat0')  edge                                   node [above] {${\hh_1\hh_2}!{\msg[b]}$} (2);       \end{tikzpicture}
      &
             \begin{tikzpicture}[mycfsm]
  \node[state]           (0)                        {$0$};
  \node[state]           (hat0)          [below right of=0, yshift=5mm]              {$\widehat{0}$};
   \node[draw=none,fill=none] (start) [above left = 0.3cm  of 0]{$\HH_2$};
  \node[state]            (1) [right of=hat0] {$1$};
  \node[state]           (hat0') [above right of=0, yshift=-5mm] {$\widehat{0}'$};
  \node[state]           (2) [right of=hat0'] {$2$};

   \path  (start) edge node {} (0) 
            (0)  edge                                   node [above] {${\hh_1\hh_2}?{\msg[b]}$} (hat0)
                  edge                                   node [above]  {${\ttv\hh_2}?{\msg[a]}$} (hat0')
             (hat0)  edge                       node [above] {${\HH_2\ttv}!{\msg[b]}$} (1)
             (hat0')  edge                                   node [above] {${\hh_2\hh_1}!{\msg[a]}$} (2);       \end{tikzpicture}
       &
     \begin{tikzpicture}[mycfsm]
  \node[state]           (0)                        {$0$};
   \node[draw=none,fill=none] (start) [above left = 0.3cm  of 0]{$\ttv$};
   \node[state]            (1) [below of=0, yshift=4mm] {$1$};

   \path  (start) edge node {} (0)
            (0)  edge    node [above] {$\hh_2\ttv?\msg[b]$} (1) ;
       \end{tikzpicture}
\end{array}
$$
The initial configuration is actually a deadlock,  and hence the system does also not enjoy progress.
\finex
\end{example}

 \begin{example}[No mixed-state counterexample for reception-error-freeness preservation]
 \label{ex:refpres}
\em
Let us consider the two following systems $S_1$ and $S_2$ with interfaces,
respectively, $\hh_1$ and $\hh_2$
containing mixed states.
$$
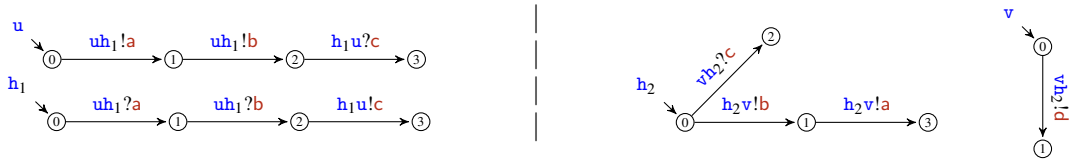
\begin{array}{c@{\hspace{1cm}}c@{\hspace{1cm}}c@{\qquad}c}
    \begin{array}{c}
      \begin{tikzpicture}[mycfsm]
  \node[state]           (0)                        {$0$};
   \node[draw=none,fill=none] (start) [above left = 0.3cm  of 0]{$\ttu$};
  \node[state]            (1) [ right of=0] {$1$};
  \node[state]            (2) [ right of=1] {$2$};
  \node[state]            (3) [ right of=2] {$3$};

   \path  (start) edge node {} (0) 
            (0)  edge                                   node [above] {$\ttu\hh_1!\msg[a]$} (1)
            (1)  edge                                   node [above] {$\ttu\hh_1!\msg[b]$} (2)
            (2)  edge                                   node [above] {$\hh_1\ttu?\msg[c]$} (3);
       \end{tikzpicture}
\\
      \begin{tikzpicture}[mycfsm]
  \node[state]           (0)                        {$0$};
   \node[draw=none,fill=none] (start) [above left = 0.3cm  of 0]{$\HH_1$};
  \node[state]            (1) [ right of=0] {$1$};
  \node[state]            (2) [ right of=1] {$2$};
  \node[state]            (3) [ right of=2] {$3$};

   \path  (start) edge node {} (0) 
            (0)  edge                                   node [above] {$\ttu\hh_1?\msg[a]$} (1)
            (1)  edge                                   node [above] {$\ttu\hh_1?\msg[b]$} (2)
            (2)  edge                                   node [above] {$\hh_1\ttu!\msg[c]$} (3);
       \end{tikzpicture}
    \end{array}
       &
       \begin{array}{c}
       |\\
       |\\
       |\\
       |
       \end{array}
       &
       \begin{tikzpicture}[mycfsm]
  \node[state]           (0)                        {$0$};
   \node[draw=none,fill=none] (start) [above left = 0.3cm  of 0]{$\hh_2$};
  \node[state]            (1) [right of=0] {$1$};
  \node[state]           (2) [above right of=0] {$2$};
    \node[state]           (3) [right of=1] {$3$};
 %       \node[state]           (five) [above right of=two] {$5$};

   \path  (start) edge node {} (0) 
            (0)  edge                                   node [above] {$\hh_2\ttv!\msg[b]$} (1)
                      edge                                   node [above]  {$\ttv\hh_2?\msg[c]$} (2)
	     (1)  edge                                   node [above] {$\hh_2\ttv!\msg[a]$} (3);
%	              edge                                   node [above] {$\hh_2\tts!\msg[d]$} (five);
       \end{tikzpicture}
&
      \begin{tikzpicture}[mycfsm]
  \node[state]           (0)                        {$0$};
   \node[draw=none,fill=none] (start) [above left = 0.3cm  of 0]{$\ttv$};
  \node[state]            (1) [ below of=0, yshift = 3mm] {$1$};

   \path  (start) edge node {} (0) 
            (0)  edge                                   node [above] {$\ttv\hh_2!\msg[d]$} (1);
       \end{tikzpicture}
\end{array}
$$

\noindent
$S_1$ and $S_2$ are both reception-error free. 
The unique communication model for their composition is\\
\centerline{$
\cm = \Set{(\hh_1,\msg[a],\hh_2),(\hh_1,\msg[b],\hh_2),(\hh_2,\msg[c],\hh_1)}
$}
The unique communication policy complying with $\cm$ is
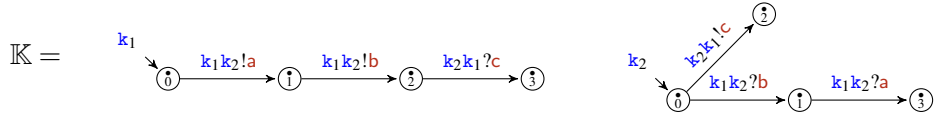
$$
\cp = \quad
\begin{array}{c@{\hspace{1cm}}c}
      \begin{tikzpicture}[mycfsm]
  \node[state]           (0)                        {$\dot 0$};
   \node[draw=none,fill=none] (start) [above left = 0.3cm  of 0]{$\kk_1$};
  \node[state]            (1) [ right of=0] {$\dot 1$};
  \node[state]            (2) [ right of=1] {$\dot 2$};
  \node[state]            (3) [ right of=2] {$\dot 3$};

   \path  (start) edge node {} (0) 
            (0)  edge                                   node [above] {$\kk_1\kk_2!\msg[a]$} (1)
            (1)  edge                                   node [above] {$\kk_1\kk_2!\msg[b]$} (2)
            (2)  edge                                   node [above] {$\kk_2\kk_1?\msg[c]$} (3);
       \end{tikzpicture}
       &
       \begin{tikzpicture}[mycfsm]
  \node[state]           (0)                        {$\dot 0$};
   \node[draw=none,fill=none] (start) [above left = 0.3cm  of 0]{$\kk_2$};
  \node[state]            (1) [right of=0] {$\dot 1$};
  \node[state]           (2) [above right of=0] {$\dot 2$};
    \node[state]           (3) [right of=1] {$\dot 3$};
 
   \path  (start) edge node {} (0) 
            (0)  edge                                   node [above] {$\kk_1\kk_2?\msg[b]$} (1)
                      edge                                   node [above]  {$\kk_2\kk_1!\msg[c]$} (2)
	     (1)  edge                                   node [above] {$\kk_1\kk_2?\msg[a]$} (3);
       \end{tikzpicture}
\end{array}
$$
Also $\cp$ is reception-error free. The system $\MC(\Set{S_1,S_2}, \cp)$  is the following one.
$$
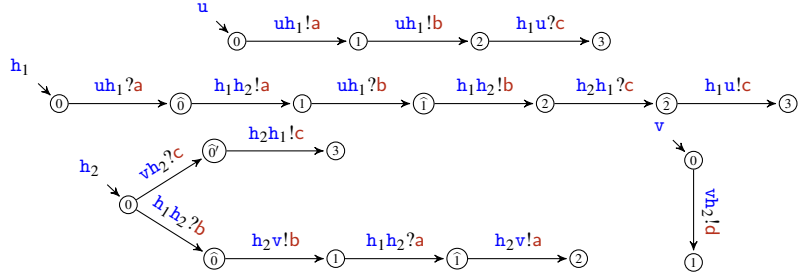
\begin{array}{c}
      \begin{tikzpicture}[mycfsm]
  \node[state]           (0)                        {$0$};
   \node[draw=none,fill=none] (start) [above left = 0.3cm  of 0]{$\ttu$};
  \node[state]            (1) [ right of=0] {$1$};
  \node[state]            (2) [ right of=1] {$2$};
  \node[state]            (3) [ right of=2] {$3$};

   \path  (start) edge node {} (0) 
            (0)  edge                                   node [above] {$\ttu\hh_1!\msg[a]$} (1)
            (1)  edge                                   node [above] {$\ttu\hh_1!\msg[b]$} (2)
            (2)  edge                                   node [above] {$\hh_1\ttu?\msg[c]$} (3);
       \end{tikzpicture}
 \\
      \begin{tikzpicture}[mycfsm]
  \node[state]           (0)                        {$0$};
  \node[state]           (hat0)          [right of=0]              {$\widehat{0}$};
   \node[draw=none,fill=none] (start) [above left = 0.3cm  of 0]{$\HH_1$};
  \node[state]            (1) [right of=hat0] {$1$};
  \node[state]           (hat1) [right of=1] {$\widehat{1}$};
  \node[state]           (2) [right of=hat1] {$2$};
      \node[state]           (hat2) [right of=2] {$\widehat{2}$};
    \node[state]           (3) [right of=hat2] {$3$};

   \path  (start) edge node {} (0) 
            (0)  edge                                   node [above] {${\ttu\hh_1}?{\msg[a]}$} (hat0)
             (1)     edge                                   node [above]  {${\ttu\hh_1}?{\msg[b]}$} (hat1)
             (hat0)  edge                       node [above] {${\HH_1\HH_2}!{\msg[a]}$} (1)
             (hat1)  edge                                   node [above] {${\hh_1\hh_2}!{\msg[b]}$} (2)
              (2)  edge                                   node [above] {${\hh_2\hh_1}?{\msg[c]}$} (hat2)
            (hat2)  edge                                   node [above]{${\hh_1\ttu}!{\msg[c]}$} (3);
       \end{tikzpicture}
       \\
             \begin{tikzpicture}[mycfsm]
  \node[state]           (0)                        {$0$};
  \node[state]           (hat0)          [below right of=0, yshift=5mm]              {$\widehat{0}$};
   \node[draw=none,fill=none] (start) [above left = 0.3cm  of 0]{$\HH_2$};
  \node[state]            (1) [right of=hat0] {$1$};
  \node[state]           (hat0') [above right of=0, yshift=-5mm] {$\widehat{0}'$};
  \node[state]           (3) [right of=hat0'] {$3$};
   \node[state]           (hat1) [ right of=1] {$\widehat{1}$};
    \node[state]           (2) [right of=hat1] {$2$};

   \path  (start) edge node {} (0) 
            (0)  edge                                   node [above] {${\hh_1\hh_2}?{\msg[b]}$} (hat0)
                  edge                                   node [above]  {${\ttv\hh_2}?{\msg[c]}$} (hat0')
             (hat0)  edge                       node [above] {${\HH_2\ttv}!{\msg[b]}$} (1)
             (hat0')  edge                                   node [above] {${\hh_2\hh_1}!{\msg[c]}$} (3)
              (1)    edge                                   node [above] {${\hh_1\hh_2}?{\msg[a]}$} (hat1)
             (hat1)  edge                                   node [above] {${\hh_2\ttv}!{\msg[a]}$} (2);
       \end{tikzpicture}
       \qquad
      \begin{tikzpicture}[mycfsm]
  \node[state]           (0)                        {$0$};
   \node[draw=none,fill=none] (start) [above left = 0.3cm  of 0]{$\ttv$};
  \node[state]            (1) [ below of=0, yshift = 3mm] {$1$};

   \path  (start) edge node {} (0) 
            (0)  edge                                   node [above] {$\ttv\hh_2!\msg[d]$} (1);
       \end{tikzpicture}
\end{array}
$$
This communication system, however, is not reception-error free, since it is possible to reach the 
configuration $s = (\vec q,\vec w)$ where\\
\centerline{
$
\vec{q} = (2_{\ttu}, 2_{\hh_1}, 0_{\hh_2}, 1_{\ttv}),
\qquad  w_{\hh_1\hh_2} = \langle\msg[a]\cdot\msg[b]\rangle,
\qquad w_{\ttv\hh_2} = \langle\msg[d]\rangle,
\qquad w_{c} = \varepsilon \ \ (\forall c\not\in\Set{\hh_1\hh_2, \ttv\hh_2})
%
%\vec{w} = (\langle\msg[a]\cdot\msg[b]\rangle_{\hh_1\hh_2}, \langle\msg[d]\rangle_{\ttv\hh_2},\vec\varepsilon)
$}
In the configuration $s$, the CFSM $\hh_2$ is in a receiving state, namely $0$, from which there are two transitions, namely
$(0,{\ttv\hh_2}?{\msg[c]},\hat 0')$ and  $(0, {\hh_1\hh_2}?{\msg[b]}, \hat 0)$.
Moreover, the channels $\ttv\hh_2$ and $\hh_1\hh_2$ are both not empty and their first element
is different from both $\msg[b]$ and $\msg[c]$.
The above configuration is hence an unspecified reception configuration.
\finex
\end{example}

Notice that 
in case we dropped the requirement that $\cp$ has to comply with a communication model,
the interfaces $\hh_1$ and $\hh_2$ of \cref{ex:refpres} could  be simplified
 to get the counterexample. In particular, they
could have just, respectively, two and three states.
The use of communication models hence %sort of limit
limits the possibility of getting systems whose properties are not preserved by composition. 
 This is an indication that connection models increase
the possibility of getting safe compositions.

%Reception-error freedom preservation is however recoverable by requiring interfaces
%to have no-mixed states, as shown in our main theorem below. 
%In fact, what causes the presence of the unspecified reception configuration in the above
%counterexample is essentially  the mixed state $0$ of $\hh_2$.

%The main goal of the present paper is the investigation of communication property preservation
%in case the properties enjoyed by the systems we compose are also enjoied by the connection
%policy we use for the multicomposition.
%This is far from being always true. 

\medskip
Let us now turn to the last communication property stated in~\cref{def:safeness}
which is lock-freeness. This property is also meaningful in the context of synchronous
communication.

%As mentioned in the Introduction, 
%In \cite[Example 6.7]{BLT23}
%a counterexample is provided, showing that in the formalism of {\em synchronous} CFSMs the properties of (synchronous) lock-freeness and deadlock-freeness
%are, in general, not preserved even in case the connection policy does. A rather strict condition is in fact required in \cite{BLT23} in order to get property preservation by composition.
%Even by considering a looser communication model (dubbed {\em asymmetric synchronous}, where
%the sender autonomously decides the message to be sent as well as
%the intended receiver) lock-freeness is not preserved, whereas the other property is.
%As a matter of fact, lock-freeness is problematic also for the case of asynchronous communications
%and no mixed states,
%as shown by the following example,
%adapted from~\cite{BLT23}.%one in \cite{BLT23}.

In \cite[Example 6.7]{BLT23}
a counterexample is provided, showing that in the formalism of {\em synchronous} CFSMs the properties of (synchronous) lock-freeness and deadlock-freeness
are, in general, not preserved.
As a matter of fact, lock-freeness is problematic also for the case of asynchronous communications and no mixed states,
as shown in the following example,
adapted from~\cite{BLT23}.%one in \cite{BLT23}.

%\footnote{As
%a matter of fact, also the property of strong lock-freeness is not preserved using symmetric synchronous
%communications, but preserved using asymmetric synchronous communications.
%We shall define and discuss the  strong lock-freeness in Section \ref{sect:conclusions}.}.

\begin{example}[Lock-freeness is not preserved by composition]\label{rem:lfnotpres} 
{\em  Let us consider the following communicating systems $S_1$ and $S_2$.\\[-4mm]
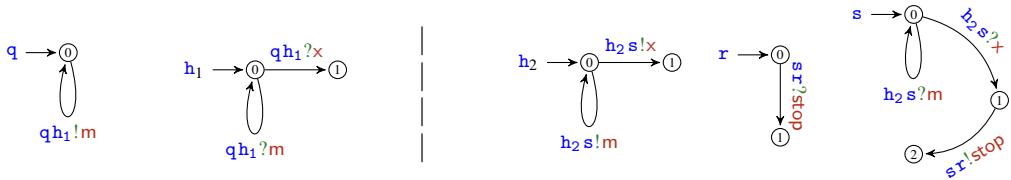
$$
  % 
%  \begin{align*}
%	S_1 = \dboxed{
	 \begin{array}{c}
		\begin{tikzpicture}[mycfsm]
		  \node[state, initial, initial where = left, initial text={$\ptp[q]$}] (zero) {$0$};
		  %\node[state] (one) [below of=zero]   {$1$};
		  % 
		  \path
		  (zero) edge [loop below,looseness=40] node[below] {$\aout[q][h_1][][m]$} (zero)
		 % (one) edge[bend left]  node[above] {$\aout[a][h][][m]$} (zero)
		  ;
		\end{tikzpicture}
	 \end{array}
	 \qquad
	 \begin{tikzpicture}[mycfsm]
		\node[state, initial, initial where = left, initial text={$\HH_1$}] (zero) {$0$};
      \node[state] (one) [right = 1cm of zero]   {$1$};
      \path
      (zero) edge [loop below,looseness=40] node[below] {$\ain[q][h_1][][m]$} (zero)
      (zero) edge node[above] {$\ain[q][h_1][][x]$} (one)
      ;
	 \end{tikzpicture}
%	 }
	 \qquad
             \begin{array}{c}
       |\\
       |\\
       |\\
       |
       \end{array}
      \qquad      
%	 S_2 = \dboxed{
	 \begin{array}{ccc}
		\begin{tikzpicture}[mycfsm]
		  \node[state, initial, initial where = left,  initial text={$\HH_2$}] (zero) {$0$};
		 % \node[state] (one) [left of=zero]   {$1$};
		  \node[state] (two) [right = 1cm of zero]   {$1$};
		  %\node[state] (three) [below = 1cm of two]   {$2$};
		  %
		  \path
		 % (zero) edge[bend right] node[above] {$\tau$} (one)
		  (zero) edge node[above] {$\aout[h_2][s][][x]$} (two)
		  %(two) edge node[above] {$\aout[h_2][s][][x]$} (three)
		  (zero) edge [loop below,looseness=40] node[below] {$\aout[h_2][s][][m]$} (zero)
		  ;
		\end{tikzpicture}
		\quad
		\begin{tikzpicture}[mycfsm]
		  \node[state, initial, initial where = left, initial text={$\ttr$}] (zero) {$0$};
		  \node[state] (one) [below = 1cm of zero]   {$1$};
		  \path (zero) edge node[above] {$\ain[s][r][][stop]$} (one)
		  ;
		\end{tikzpicture}
		\quad
		\begin{tikzpicture}[mycfsm]
		  \node[state, initial, initial where = left, initial text={$\ptp[s]$}] (zero) {$0$};
		  \node[state] (one) [below right of=zero]   {$1$};
		  \node[state] (two) [below left of=one, yshift=.5cm]   {$2$};
		  %\node[state] (three) [above left of=two]   {$3$};
		  % 
		  \path
		  (zero) edge[loop below,looseness=40] node[below] {$\ain[h_2][s][][m]$} (zero)
		  (zero) edge [bend left] node[above] {$\ain[h_2][s][][x]$} (one)
		  (one) edge [bend left]  node[below] {$\aout[s][r][][stop]$} (two)
		 % (two) edge [bend left]  node[below] {$\aout[s][r][][stop]$} (three)
		  ;
		\end{tikzpicture}
	 \end{array}
$$
%	 }
%  \end{align*}
  Note that both $S_1$ and $S_2$ are lock-free and their respective interfaces $\HH_1$ and $\HH_2$ have no mixed states.
% $\Set{S_i}_{i=1,2}$ is composable with respect 
  %to the set of interfaces 
   
   Let us now consider the  (unique) connection policy % for $H$:\quad
   %complying with the obvious communication model
   $\cp =(M_{\kk_i})_{i\in \Set{1,2}}$
   where $M_{\kk_1}\in \IS {M_{\HH_1}}{\cm}$
   and $M_{\kk_2}\in \IS {M_{\HH_2}}{\cm}$ 
   with connection model $\cm = \Set{(\hh_1,\msg[x],\hh_2),(\hh_1,\msg[m],\hh_2)}$.\\ 
   %In particular,\\
%   \begin{align*}
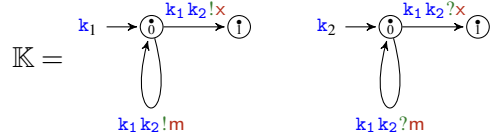
$$
	\cp = %\dboxed{
	 \begin{tikzpicture}[mycfsm]
		\node[state, initial, initial where = left, initial text={$\KK_1$}] (zero) {$\dot 0$};
      \node[state] (one) [right = 1cm of zero]   {$\dot 1$};
      \path
      (zero) edge [loop below,looseness=40] node[below] {$\aout[k_1][k_2][][m]$} (zero)
      (zero) edge node[above] {$\aout[k_1][k_2][][x]$} (one)
      ;
	 \end{tikzpicture}
	 \qquad
	 \begin{tikzpicture}[mycfsm]
		  \node[state, initial, initial where = left, initial text={$\KK_2$}] (zero) {$\dot 0$};
		 % \node[state] (one) [left of=zero]   {$1$};
		  \node[state] (two) [right = 1cm of zero]   {$\dot 1$};
		  %\node[state] (three) [below = 1cm of two]   {$2$};
		  %
		  \path
		 % (zero) edge[bend right] node[above] {$\tau$} (one)
		  (zero) edge node[above] {$\ain[k_1][k_2][][x]$} (two)
		  %(two) edge node[above] {$\aout[h_2][s][][x]$} (three)
		  (zero) edge [loop below,looseness=40] node[below] {$\ain[k_1][k_2][][m]$} (zero)
		  ;
	    \end{tikzpicture}
%	                        }
%         \end{align*}
$$
   %
%   Note that, since we are considering a connection policy for two systems only, both $\IS {M_{\HH_1}}{\cm}$ and $\IS {M_{\HH_2}}{\cm}$ are necessarily singletons.
   It is easy to see that $\cp$ is lock-free.
   The multicomposition  $\MC(\Set{S_i}_{i\in \Set{1,2}}, \cp)$ is the following communicating system: 
$$ 
%  \begin{align*}
%     \dboxed{
       
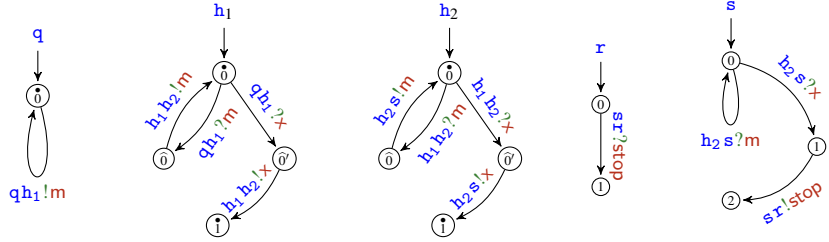
\begin{tikzpicture}[mycfsm]
		  \node[state, initial, initial where = above, initial text={$\ptp[q]$}] (zero) {$\dot 0$};
		  \path
		  (zero) edge [loop below,looseness=40] node[below] {$\aout[q][h_1][][m]$} (zero)
		 % (one) edge[bend left]  node[above] {$\aout[a][h][][m]$} (zero)
		  ;
		\end{tikzpicture}
		\qquad
	 \begin{tikzpicture}[mycfsm]
		\node[state, initial, initial where = above, initial text={$\HH_1$}] (zero) {$\dot 0$};
           \node[state] (zerohat) [below left of=zero,xshift=4mm]   {$\widehat 0$};
	     \node[state] (zerohatp) [below right of=zero,xshift=-4mm]   {$\widehat 0'$};	
	      \node[state] (one) [below left of=zerohatp, xshift=3mm,yshift=3mm]   {$\dot 1$};  
      \path
      (zero) edge [bend left=20] node[below] {$\ain[q][h_1][][m]$} (zerohat)
      (zerohat) edge [bend left=20] node[above] {$\aout[h_1][h_2][][m]$} (zero)
      (zero) edge node[above] {$\ain[q][h_1][][x]$} (zerohatp)
      (zerohatp) edge [bend left=20]  node[above] {$\aout[h_1][h_2][][x]$} (one)
      ;
	 \end{tikzpicture}
	 \qquad
	 \begin{tikzpicture}[mycfsm]
		  \node[state, initial, initial text={$\HH_2$}] (zero) {$\dot 0$};
           \node[state] (zerohat) [below left of=zero,xshift=4mm]   {$\widehat 0$};
	     \node[state] (zerohatp) [below right of=zero,xshift=-4mm]   {$\widehat 0'$};	
	      \node[state] (one) [below left of=zerohatp, xshift=3mm,yshift=3mm]   {$\dot 1$};  
		  \path
		 % (zero) edge[bend right] node[above] {$\tau$} (one)
		  (zero) edge node[above] {$\ain[h_1][h_2][][x]$} (zerohatp)
		  (zerohatp) edge [bend left=20] node[above] {$\aout[h_2][s][][x]$} (one)
		  %(two) edge node[above] {$\aout[h_2][s][][x]$} (three)
		  (zero) edge [bend left=20] node[below] {$\ain[h_1][h_2][][m]$} (zerohat)
		  (zerohat) edge [bend left=20] node[above] {$\aout[h_2][s][][m]$} (zero)
		  ;
	    \end{tikzpicture}
	    \qquad
	    \begin{tikzpicture}[mycfsm]
		  \node[state, initial, initial where = above, initial text={$\ttr$}] (zero) {$0$};
		  \node[state] (one) [below = 1cm of zero]   {$1$};
		  \path (zero) edge node[above] {$\ain[s][r][][stop]$} (one)
		  ;
		\end{tikzpicture}
		\qquad
		\begin{tikzpicture}[mycfsm]
		  \node[state, initial, initial where = above, initial text={$\ptp[s]$}] (zero) {$0$};
		  \node[state] (one) [below right of=zero]   {$1$};
		  \node[state] (two) [below left of=one, yshift=.5cm]   {$2$};
		  %\node[state] (three) [above left of=two]   {$3$};
		  % 
		  \path
		  (zero) edge[loop below,looseness=40] node[below] {$\ain[h_2][s][][m]$} (zero)
		  (zero) edge [bend left] node[above] {$\ain[h_2][s][][x]$} (one)
		  (one) edge [bend left]  node[below] {$\aout[s][r][][stop]$} (two)
		  %(two) edge [bend left]  node[below] {$\aout[s][r][][stop]$} (three)
		  ;
		\end{tikzpicture}
%	                        }
  $$
  %     \end{align*} 
  % 
  The initial configuration $s_0$ of $\MC(\Set{S_i}_{i\in \Set{1,2}}, \cp)$ is an
  $\ptp[r]$-lock,  since the transition $\ain[q][h_1][][x]$ of $\HH_1$ can never be fired,
  so implying, in turn, that also $\aout[h_1][h_2][][x]$ of $\HH_1$,
  $\ain[h_1][h_2][][x]$ of $\HH_2$, $\aout[h_2][s][][x]$ of $\HH_2$,
  $\ain[h_2][s][][x]$ of $\tts$ and $\aout[s][r][][stop]$ of $\tts$ can never be fired.
  Hence, no transition sequence out of $s_0$ will ever involve the participant  $\ttr$.
  Thus $\MC(\Set{S_i}_{i\in \Set{1,2}}, \cp)$ is not lock-free. 
  }
  \finex
\end{example} 

\smallskip
\begin{remark}
{\em
It is worth noticing that, in Examples \ref{ex:lackprogdfpres}, \ref{ex:refpres} and \ref{rem:lfnotpres}
above, the interfaces of the systems we compose do have unreachable states.
It is hence natural to wonder whether it is the presence of unreachable states in interfaces that entails the possibility of getting counterexamples for the properties taken into account. 
}\finex
\end{remark}

%!TEX root = Main-asynchCFSM-multicomp.tex

\section{Conclusions}
\label{sect:conclusions}

The necessity of supporting the modular development of concurrent/distributed systems,
as well as the need to extend/modify/adapt/upgrade them, urged the investigation
 of composition methods. Focusing on such investigations in the setting of abstract formalisms
for the description and verification of systems enables to get general and formal guarantees of relevant features
of the composition methods. 

An investigation of composition in a formalism for choreographic programming was carried out in \cite{MY13}. 
In \cite{KFG04} a modular technique was developed for the verification of 
aspect-oriented programs expressed as state machines.
Team Automata is another formalism in which compositionality issues have been addressed
\cite{BK03,BHK-ictac20},
as well as in assembly theories considered in~\cite{HK-acta15}. 
Composition for protocols described via a process algebra has been investigated in \cite{BOV23}.
In \cite{CMV18,SGV20} a technique for modular design in the setting of reactive programming
is proposed. A possible approach to composition for a MultiParty Session Type (MPST) formalism
is developed in \cite{SMG23}. 
The mentioned papers provide just a glimpse of the variety of approaches to system composition in the literature.

%In~\cite{BdLH19} a quite general approach to (binary) composition of systems
%(dubbed {\em participants-as-interfaces\/} -- PaI -- in subsequent papers) was introduced
%and exploited for the asynchronous formalism of Communicating Finite State Machines (CFSM).
%It was also investigated in \cite{BLT20,BLP22b,BLT23}, always for binary composition, for 
%a synchronous version of such a formalism, as well in~\cite{BDL22,BDLT21} for synchronous MPTS
%formalisms.
%The PaI approach distils to the interpretation of participants as interfaces 
%and to their replacement with forwarders for the composition of systems. 
%The extension of such an approach in order to compose more than two systems
%was investigated in a synchronous MPTS formalism in \cite{BDGY23}.

Papers dealing with the (binary) composition of systems on the basis of the
{\em participants-as-interfaces\/} (PaI) approach have been pointed out already in~\cref{sec:Intro} and the idea of PaI for multicomposition of systems has been explained in~\cref{sec:pai-multicomp}.
In the present paper we  study  the PaI approach to multicomposition
 for systems of asynchronously communicating finite state machines (CFSMs).
We show that  (under mild assumptions) important 
communication properties relevant  in the context of asynchronous communication, like freeness of orphan messages and unspecified receptions,  are preserved by composition (a feature dubbed 
{\em safety\/} in \cite{BDGY23}).
 For this we assume that 
for each single system one participant is chosen as an interface. 
A key role in our work, inspired by  \cite{BDGY23}, is played by
{\em connection policies}, which are CFSM systems which determine the ways how interfaces can interact
when they are replaced by gateways (forwarders) in system compositions.

For an ``unstructured'' formalism like CFSM,
the natural generalisation from multicomposition with
single interfaces to multicomposition with multiple interfaces  (per system)  is not trouble-free,
as discussed in \cite[Sect.6]{BdLH19} for binary composition.
This is mainly due to the possible indirect  interactions  which could occur  among the interfaces inside the single systems.
In more structured formalisms, however, such possible interactions can be controlled. 
This is the case, for instance, in MPST formalisms.
In fact, in \cite{GY23} the authors devise a  
direct composition mechanism without using gateways for MPST systems.
Such a mechanism allows for the presence of multiple interfaces thanks to an hybridisation with local and external information of the standard notion of global type.  
A combination of global and local constructs in order to get flexible specifications
(uniformly describing both the internal and the interface behavior of systems) is also present in \cite{CV10}.

There are several directions to be pursued in future work starting from our results.
On the first place, we want to generalise the notion of connection policy such
that PaI  multicomposition could actually be obtained by replacing interfaces
by gateways which, instead of interacting directly with each other, can interact  through an ``interfacing infrastructure'' represented via a system of CFSMs. Such a generalisation would be equivalent to multicomposition
where exactly one system can have multiple interfaces.
 Let us consider a possible application of the above idea. 
In \cref{ex:simplewe}, in the resulting composed system, both participants $\ttp$ and $\ttq$ 
do emit a $\msg[react]$ message. 
It would be more natural to have only one of them producing such a message, e.g., to have $\ttp$ be the sole sensor registering reactions which then passes that information to both $\ttr$ and $\tts$.
This would not be possible by our composition mechanism and
we cannot but make the best of the fact that we are dealing with two sensors.
One could think, instead, about using an “interfacing infrastructure”  containing some further participant enabling to ignore the messages from one sensor and properly duplicating the messages from the other.

%\brc
%Here I am wondering whether our goal shouldn't be to get
%multicomposition where each system is enabled to have multiple interfaces
%?
%Then I would rewrite the last part to:
%\erc
%
%\brr
%There are several directions to be pursued in future work starting from our results.
%On the first place, we want to generalise the notion of connection policy such
%that PaI  multicomposition could be based on composition with
%several interfaces per system following ideas of the hybrid approach above
%but using asynchronous communication and CFSM systems. 
%\err

%\bfr
%In the composition approach exploited in the present paper, every communication of an interface must be received/forwarded by the correponding gateway from/to some other gateway. It is reasonable to generalise this approach such that only  certain messages are received/forwarded, while for others the interface keeps aquiring/producing them by itself. Such an idea was actually implemented in \cite{BDL22} in a MPTS setting for a restricted client-multiserver composition with synchronous communications.
%

%It is worth noticing that, in Examples \ref{ex:lackprogdfpres}, \ref{ex:refpres} and \ref{rem:lfnotpres},
%the interfaces of the systems we compose do have unreachable states.
%It is hence natural to wonder whether it is the presence of unreachable states in interfaces that entails the possibility of getting counterexamples for the properties taken into account. 

We are also planning to consider further communication properties, like strong lock-freeness
(any participant can eventually progress in any continuation of any reachable configuration),
as well as to investigate conditions to get lock-freeness preservation, not guaranteed yet.

Unlike the present paper, in \cite{BdLH19} safety is ensured for the binary case by assuming 
compatibility of interfaces and an extra condition (called ?!-determinism) on them.
We are currently considering a generalisation of the binary compatibility relation.
 Such generalisation should imply relevant communication properties for the communication policy it depends on.

% We are also planning to identify some conditions  ensuring
%\cref{th:paisafenesse} to hold for any communication property $\mathcal{P}$ satisfying them.

 Finally, we are interested in considering “partial” gateways, where only some communications of an interface are interpreted as communications with the environment.
Such an idea was actually implemented in \cite{BDL22} in a MPTS setting for a restricted client-multiserver composition with synchronous communications.
% Such an extension can be worth investigating, as partially done, in the context of MPTS with synchronous communications, in \cite{BDL22}.

%\paragraph{Acknowledgements}
%We are grateful to the anonymous referees for several helpful comments and suggestions.
%We also thank Emilio Tuosto for some macros used to draw Figure \ref{fig:examplegg}.
%The first author is also thankful to Mariangiola Dezani for her everlasting support.
\paragraph*{\bf Acknowledgements}
 We warmly thank the ICE'24 reviewers for their careful reading, their thoughtful comments/suggestions and the helpful discussion in the forum. We also thank Emilio Tuosto for his nice tikz style for automata.
\setlength{\abovedisplayskip}{6pt}
\setlength{\belowdisplayskip}{\abovedisplayskip}

\bibliographystyle{eptcs}
\bibliography{session}

%------APPENDIX--------
%\newpage
%\appendix
%
%\input{safenessPreservation}
%\input{weakDeadlockPreservation}
%\input{strong-deadlock-free-Preservation}
%---------------------------

%------------------------------------------------------------------------------
% Index
%\printindex

%------------------------------------------------------------------------------
\end{document}

% EOF